\documentclass[preprint,12pt]{elsarticle}
\usepackage{graphicx}
\usepackage{amssymb}
\usepackage{ulem}
\usepackage{here}
\journal{Nuclear Physics A}

\begin{document}
\begin{frontmatter}
\title{Nuclear equation of state for core-collapse supernova simulations with realistic nuclear forces}

\author[RIKEN,RISE]{H.~Togashi\corref{cor1}}
\ead{hajime.togashi@riken.jp}
\author[Kyushu]{K.~Nakazato}
\author[TUS]{Y.~Takehara}
\author[TUS]{S.~Yamamuro}
\author[TUS]{H.~Suzuki}
\author[RISE,WU]{M.~Takano}

\cortext[cor1]{Corresponding author}

\address[RIKEN]{Nishina Center for Accelerator-Based Science, Institute of Physical and Chemical Research (RIKEN), 
2-1 Hirosawa, Wako, Saitama 351-0198, Japan}
\address[RISE]{Research Institute for Science and Engineering, 
Waseda University, 3-4-1 Okubo Shinjuku-ku, Tokyo 169-8555, Japan}
\address[Kyushu]{Faculty of Arts and Science, 
Kyushu University, 744 Motooka, Nishi-ku, Fukuoka 819-0395, Japan}
\address[TUS]{Department of Physics, Faculty of Science and Technology, 
Tokyo University of Science, Yamazaki 2641, Noda, Chiba 278-8510, Japan}
\address[WU]{Department of Pure and Applied Physics, 
Graduate School of Advanced Science and Engineering, 
Waseda University, 3-4-1 Okubo Shinjuku-ku, Tokyo 169-8555, Japan}

\begin{abstract}
A new table of the nuclear equation of state (EOS) based on realistic nuclear potentials is constructed for core-collapse supernova numerical simulations.  
Adopting the EOS of uniform nuclear matter constructed by two of the present authors with the cluster variational method starting from the Argonne v18 and Urbana IX nuclear potentials, 
the Thomas-Fermi calculation is performed to obtain the minimized free energy of a Wigner-Seitz cell in non-uniform nuclear matter.  
As a preparation for the Thomas-Fermi calculation, the EOS of uniform nuclear matter is modified so as to remove the effects of deuteron cluster formation in uniform matter at low densities.
Mixing of alpha particles is also taken into account following the procedure used by Shen et al. (1998, 2011).  
The critical densities with respect to the phase transition from non-uniform to uniform phase with the present EOS are slightly higher than those with the Shen EOS at small proton fractions.  
The critical temperature with respect to the liquid-gas phase transition decreases with the proton fraction in a more gradual manner than in the Shen EOS.  
Furthermore, the mass and proton numbers of nuclides appearing in non-uniform nuclear matter with small proton fractions are larger than those of the Shen EOS.  
These results are consequences of the fact that the density derivative coefficient of the symmetry energy of our EOS is smaller than that of the Shen EOS.  
\end{abstract}


\end{frontmatter}
\section{Introduction}
Core-collapse supernova explosions occur at the end of the evolution of massive stars heavier than 8 $M_{\odot}$.  
At the final stage of these stars, the stellar central core is composed of highly-degenerate electrons and heavy nuclides such as irons.  
Owing to photodisintegration of heavy nuclei and electron capture reactions, the core becomes gravitationally unstable, and begins to collapse. 
During the collapse, neutrinos created through the weak interaction are trapped in the core because they are scattered by other particles, mainly heavy nuclei.
When the central density of the core approaches the nuclear saturation density, 
the core becomes stiffer and the resulting core bounce produces an outgoing shock wave. 
The shock wave stalls once as a result of the energy loss caused by the photodisintegration of in-falling heavy nuclei and neutrino emissions.  
According to the neutrino-heating mechanism, a convincing scenario for core-collapse supernovae~\cite{SNreview}, 
the stalled shock then revives as neutrinos emitted from inside the core react with matter through the weak interaction and deposit energies to matter behind the shock.  

The above-described scenario of the core-collapse supernovae relies on two characteristics of nuclear matter: 
The stiffness of high-density nuclear matter, which causes a bounce of the core, and the species of nuclides in hot matter with which the neutrinos react.  
These two characters are given in the nuclear equation of state (EOS).  
In particular, because the stiffness of nuclear matter is governed by the repulsion of nuclear forces, 
it should be described with a nuclear Hamiltonian composed of realistic nuclear potentials.  
However, in the supernova EOSs proposed so far, the relation between bare nuclear forces and stiffness of nuclear matter is unclear as those EOSs are based on phenomenological nuclear models.  
The Lattimer-Swesty EOS~\cite{LS}, which is one of the standard supernova EOSs, is constructed with an effective Skyrme interaction, 
while the Shen EOS~\cite{Shen1, Shen2, Shen3}, which is another standard supernova EOS, is constructed based on the relativistic mean field (RMF) theory with the parameter set TM1~\cite{TM1}.  
The Shen EOS was extended so as to take into account hyperon mixing~\cite{Shen3, Ishizuka} and quark-hadron phase transition~\cite{Nakazato}.  
While the symmetry energy of the Shen EOS (36.9 MeV) is rather larger than the empirical value~\cite{expS}, 
supernova EOSs based on RMF theories other than TM1 have also been proposed~\cite{Hempel}. 
However, there is no supernova EOS based on bare nuclear forces yet.  

The current status of supernova EOS studies motivated us to construct a new supernova EOS based on the variational many-body theory with realistic nuclear forces.  
In this research project, we employ the Argonne v18 (AV18) two-body nuclear potential~\cite{AV18}, 
which reproduces the experimental two-nucleon scattering data and deuteron properties, 
and the Urbana IX (UIX) potential~\cite{UIX} for the three-body nuclear force.  
We then calculate the free energy per nucleon of uniform nuclear matter with the cluster variational method using the Jastrow wave function.  
Here we note that the thermodynamic quantities of asymmetric nuclear matter with arbitrary proton fractions are necessary to complete a supernova EOS table, 
while it is difficult to perform sophisticated variational calculations such as the Fermi Hypernetted Chain (FHNC) calculations for asymmetric nuclear matter.  
Therefore, we employ a simplified two-body cluster approximation and impose appropriate constraints on variational functions 
so that the obtained energies of symmetric nuclear matter and neutron matter are in good agreement with those based on the FHNC calculations. 
The EOS of uniform nuclear matter calculated in this manner is reported in Refs.~\cite{K1, K2, T1}.  
Here we note that mass-radius relation of neutron stars calculated with the obtained EOS is consistent with the observational data analyzed by Steiner et al.~\cite{Steiner}.  

As reported in Ref.~\cite{T2}, we previously applied the EOS of uniform nuclear matter in Refs.~\cite{K1,K2,T1} to numerical simulations of core-collapse supernovae.  
In that paper, we adopted the Shen EOS for non-uniform nuclear matter and connected it to our uniform nuclear EOS.  
We then performed fully general relativistic, spherically symmetric core-collapse simulations with and without neutrino transfer.  
In the case without neutrino transfer, the star exploded because the energy loss due to the neutrino emission was switched off.  
In contrast, with neutrino transfer the shock wave stalled and the stellar explosion failed. 
These results are consistent with other realistic core collapse simulations with other EOSs, 
which suggests that spherically symmetric core collapse supernova explosions are unlikely~\cite{SNreview}. 
Furthermore, the central density of our stellar core following the bounce is higher than in the case of the Shen EOS because our EOS is softer, 
i.e., the present EOS is advantageous for supernova explosions~\cite{BCK, TS, Bethe}.

Needless to say, the EOS of the non-uniform phase should be consistent with that of the uniform phase, 
i.e., it is necessary to construct an EOS of non-uniform matter in a self-consistent manner, which is the aim of this paper. 
In our project to construct a new supernova EOS, we employ the prescription for the calculation of non-uniform nuclear EOS used by Shen et al., 
i.e., we calculate nuclides in a Wigner-Seitz cell in the Thomas-Fermi approximation.  
Though several other Thomas-Fermi calculations for hot nuclear matter have also been adopted especially for the study of the pasta phase \cite{TFhot1, TFhot2, TFhot3}, 
we employ the Thomas-Fermi calculation originally adopted by Shen et al. as one of the standard Thomas-Fermi calculations for supernova EOS. 

This paper is organized as follows: 
In Section 2, we review the EOS of uniform nuclear matter adopted in this paper.
We also consider the mixing of alpha particles.  
In Section 3, we perform the Thomas-Fermi calculation in order to obtain the thermodynamic quantities of non-uniform nuclear matter following the procedure used by Shen et al.  
In Section 4, the obtained thermodynamic quantities of our EOS are compared with those in the Shen EOS.  
A summary is given in Section 5.  

\section{Equation of state for uniform nuclear matter}
\subsection{Variational method for uniform nuclear matter}
In this subsection, we describe the variational method for uniform nuclear matter that we previously developed in Ref.~\cite{T1}. 
We start from the nuclear Hamiltonian $H$ which is the sum of the two-body Hamiltonian $H_2$ and three-body Hamiltonian $H_3$.  
The two-body Hamiltonian is given as follows: 
\begin{equation}
H_2 = -{\textstyle\sum\limits_{i=1}^N} \frac{\hbar^2}{2m_{\mathrm{n}}}\nabla_i^2 + {\textstyle\sum\limits_{i<j}^N}V_{ij}, 
\label{H2}
\end{equation}
where $m_{\mathrm{n}}$ is the rest mass of a neutron. 
Note that we neglect the difference between the neutron mass and the proton mass in the kinetic energy part of the Hamiltonian, as in the previous studies~\cite{K1, K2, T1}. 
The isoscalar part (the charge-independent part) of the AV18 potential is adopted as the two-body nuclear potential $V_{ij}$. 
The three-body part of the Hamiltonian $H_3$ is constructed with the UIX three-body nuclear potential.  

At zero temperature, we assume the Jastrow wave function as follows: 
\begin{equation}
\mathnormal{\Psi}=\mathrm{Sym}\left[\prod_{i<j}f_{ij}\right]\mathnormal{\Phi}_\mathrm{F}\left[n_{0\mathrm{p}}(k), n_{0\mathrm{n}}(k)\right],   \label{jwf}
\end{equation}
where $\mathnormal{\Phi}_{\mathrm{F}}\left[n_{0\mathrm{p}}(k), n_{0\mathrm{n}}(k)\right]$ 
is the Fermi gas wave function specified by the occupation probabilities of the single-nucleon states $n_{0l}(k)=\theta\left(k_{\mathrm{F}l}-k\right)$ ($l=$ p, n), 
with $k_{\mathrm{F}l}$ being the Fermi wave numbers, and $f_{ij}$ is the correlation function between ($i, j$) nucleons.  
In addition, the operator $\mathrm{Sym}[$ $]$ in Eq.(\ref{jwf}) is the symmetrizer.  
The explicit form of the correlation function is written as
\begin{equation}
f_{ij}={\textstyle\sum\limits_{t=0}^1}{\textstyle\sum\limits_{\mu}}{\textstyle\sum\limits_{s=0}^1}
        \left[f_{\mathrm{C}ts}^{\mu}(r_{ij})+sf_{\mathrm{T}t}^{\mu}(r_{ij})S_{\mathrm{T}ij}
        +sf_{\mathrm{SO}t}^{\mu}(r_{ij})
        (\mbox{\boldmath$L$}_{ij}\cdot\mbox{\boldmath$s$})\right]P_{tsij}^{\mu}, 
\label{fij}
\end{equation}
where $f^{\mu}_{\mathrm{C}ts}(r)$, $f^{\mu}_{\mathrm{T}t}(r)$, and $f^{\mu}_{\mathrm{SO}t}(r)$ are the central, tensor, and spin-orbit correlation functions, respectively.  
These correlation functions depend on the two-nucleon total spin $s$, the total isospin $t$ and its third component expressed with $\mu$.   
Moreover, $P^{\mu}_{tsij}$ is the spin-isospin projection operator projecting a two nucleon state onto an eigenstate of $(t, s, \mu)$.  

With use of the Jastrow wave function $\mathnormal{\Psi}$ in Eq.~(\ref{jwf}), the expectation value of the two-body Hamiltonian per nucleon $\left<H_2\right>/N$ is calculated in the two-body cluster approximation.  
Since the energy contribution from the omitted higher-order cluster terms is not negligible, we also impose the following two constraints on the correlation functions $f^{\mu}_{\mathrm{C}ts}(r)$, $f^{\mu}_{\mathrm{T}t}(r)$, and $f^{\mu}_{\mathrm{SO}t}(r)$ in order to compensate for the omission of those terms~\cite{K1, T1}. 
The first one is the extended Mayer's condition which is expressed as follows: 
\begin{equation}
n_{\mathrm{B}}\int\left[F_{ts}^{\mu}(r)-F_{\mathrm{F}ts}^{\mu}(r)\right]d\mbox{\boldmath$r$}=0,  
\label{eq:mayer}
\end{equation}
where $n_{\mathrm{B}}$ is the nucleon number density given as the sum of the neutron number density $n_{\mathrm{n}}$ and the proton number density $n_{\mathrm{p}}$.  
Furthermore, $F^{\mu}_{ts}(r)$ is the $(t, s, \mu)$-projected radial distribution function expressed in the two-body cluster approximation, 
and $F^{\mu}_{\mathrm{F}ts}(r)$ is $F^{\mu}_{ts}(r)$ in the case of the Fermi gas.   
The second constraint is the healing distance condition.  
We introduce the healing distance $r_{\mathrm{h}}$ and impose $f^{\mu}_{\mathrm{C}ts}(r)=1$, $f^{\mu}_{\mathrm{T}t}(r)=f^{\mu}_{\mathrm{SO}t}(r)=0$ for $r \geq r_{\mathrm{h}}$; 
the correlation between two nucleons vanishes when the distance between them $r$ is larger than $r_{\mathrm{h}}$.  
The healing distance $r_{\mathrm{h}}$ is assumed to be independent of ($t$, $s$, $\mu$), 
and proportional to $r_{\mathrm{0}}$ which is the radius of a sphere whose volume is $1/n_{\mathrm{B}}$; i.e., $r_{\mathrm{h}}=a_{\mathrm{h}} r_0$.   
Here, the coefficient is chosen to be $a_{\mathrm{h}}=1.76$ so that the calculated two-body energies per nucleon $E_2(n_{\mathrm{p}}, n_{\mathrm{n}})$ of pure neutron matter ($n_{\mathrm{p}}=0$) and symmetric nuclear matter ($n_{\mathrm{n}}=n_{\mathrm{p}}=n_{\mathrm{B}}/2$) reproduce the results obtained by Akmal, Pandharipande, and Ravenhall (APR)~\cite{APR} with the FHNC calculation.  

In addition to the two-body Hamiltonian, we consider the energy per nucleon caused by the three-body nuclear force $E_3(n_{\mathrm{p}}, n_{\mathrm{n}})$.  
Since the UIX three-body nuclear potential consists of a phenomenological repulsive component and a two-pion exchange component, 
the corresponding three-body Hamiltonian $H_3$ is divided into the repulsive part $H^{\mathrm{R}}_3$ and the two-pion exchange part $H^{2\pi}_3$.  
For these parts of the Hamiltonian, we calculate the expectation values per nucleon with the Fermi gas wave function $\left<H_3^i\right>_{\mathrm{F}}(n_{\mathrm{p}}, n_{\mathrm{n}})/N$ ($i=\mathrm{R},2\pi$). 
Then the three-body energy per nucleon is expressed as $E_3(n_{\mathrm{p}}, n_{\mathrm{n}})=\sum_{i}\left<\alpha_iH_3^i\right>_{\mathrm{F}}(n_{\mathrm{p}}, n_{\mathrm{n}})/N+E_{\mathrm{corr}}(n_{\mathrm{p}}, n_{\mathrm{n}})$, where $\alpha_i$ are parameters representing the medium effect caused by the correlations among nucleons, and $E_{\mathrm{corr}}(n_{\mathrm{p}}, n_{\mathrm{n}})$ is an additional correction term.  
The correction term is explicitly expressed in terms of $n_{\mathrm{n}}$ and $n_{\mathrm{p}}$ with parameters, 
and slightly modifies the total energy per nucleon of nuclear matter $E(n_{\mathrm{p}}, n_{\mathrm{n}})=E_2(n_{\mathrm{p}}, n_{\mathrm{n}})+E_3(n_{\mathrm{p}}, n_{\mathrm{n}})$. 
We then determine the parameters included in $E_3(n_{\mathrm{p}}, n_{\mathrm{n}})$ so as to reproduce the empirical saturation point.  
Furthermore, these parameters are fine-tuned so that the Thomas-Fermi calculations of isolated atomic nuclei with the present $E(n_{\mathrm{p}}, n_{\mathrm{n}})$ 
reproduce the gross features of their empirical masses and radii~\cite{K2}. 
The obtained saturation density $n_0$, saturation energy $E_0$, incompressibility $K$ and symmetry energy $E_{\mathrm{sym}}$ are 
$n_0=$0.16 fm$^{-3}$, $E_0=-16.09$ MeV, $K=245$ MeV, $E_{\mathrm{sym}}=30.0$ MeV, respectively. 

\begin{figure}[t]
  \centering
  \includegraphics[width=8.0cm]{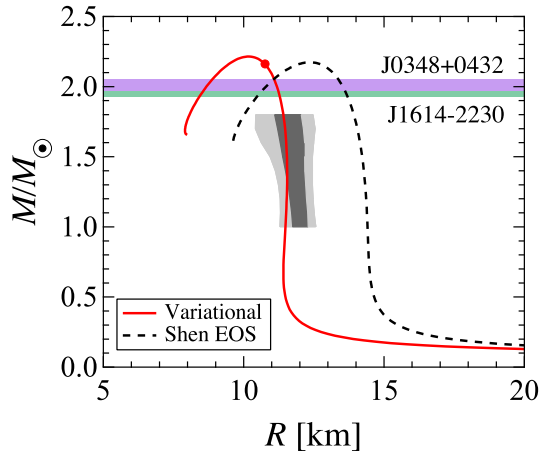}
  \caption{(Color online) Mass-radius relation of neutron stars calculated with the present EOS. 
The filled circle represents the neutron star for which the central density is equal to the critical density $n_\mathrm{c}$. 
The horizontal green and purple bands show the masses of PSRs J1614-2230 \cite{2Msolar1} and J 0348+0432 \cite{2Msolar2}. 
The shaded region indicates the mass-radius region suggested in Ref.~\cite{Steiner}. 
The mass-radius relation with the Shen EOS is also shown.} 
\label{fig:NS}
\end{figure}

The obtained $E(n_{\mathrm{p}}=n_{\mathrm{B}}/2,n_{\mathrm{n}}=n_{\mathrm{B}}/2)$ for symmetric nuclear matter 
and $E(n_{\mathrm{p}}=0,n_{\mathrm{n}}=n_{\mathrm{B}})$ for neutron matter are in good agreement with those calculated with the FHNC technique by APR.  
We also apply our $E(n_{\mathrm{p}},n_{\mathrm{n}})$ for asymmetric nuclear matter to the calculation of neutron stars. 
For the crust region, we adopt the low-density EOS that is constructed with the Thomas-Fermi method as reported below. 
The obtained maximum mass of neutron stars is found to be 2.21$M_{\odot}$, 
which is consistent with observational masses of heavy neutron stars J1614-2230 ($1.97\pm0.04M_{\odot}$)~\cite{2Msolar1} and J0348+0432 ($2.01\pm0.04M_{\odot}$)~\cite{2Msolar2}, 
as seen in Fig.~\ref{fig:NS}. 
Furthermore, the calculated mass-radius relation is in good agreement with the observationally suggested mass-radius region analyzed by Steiner et al.~\cite{Steiner}. 
It is also seen in Fig.~\ref{fig:NS} that, for a fixed gravitational mass of a neutron star, the radius with our EOS is smaller than that with the Shen EOS. 
This fact implies that our EOS is softer than the Shen EOS in the case of cold neutron star matter.  
It should be noted that the present EOS violates the causality condition at densities higher than the critical density $n_{\mathrm{c}} = 0.90$ fm$^{-3}$. 
The mass of the neutron star with the central density being $n_{\mathrm{c}}$ is 2.16$M_{\odot}$, which is also consistent with the observational data. 

At finite temperatures, we calculate the free energy per nucleon $F_{\mathrm{N}}(n_{\mathrm{p}}, n_{\mathrm{n}},T)$ at temperature $T$ 
using the prescription proposed by Schmidt and Pandharipande~\cite{SP}. 
In this method, which is based on the Landau's Fermi liquid theory, 
the entropy per nucleon $S_{\mathrm{N}}(n_{\mathrm{p}},n_{\mathrm{n}},T)$ is expressed as in the case of the non-interacting quasi-nucleon system, 
where the average occupation probability of the single nucleon state $n_{\mathrm{T}i}(k)$ ($i=$p, n) is expressed as 
\begin{equation}
n_{\mathrm{T}i}(k) = \Bigg\{1+\exp \left[\frac{\varepsilon_i(k)-\mu_{0i}}{k_{\mathrm{B}}T}\right]\Bigg\}^{-1}, 
\label{eq:nk}
\end{equation}
where $\varepsilon_i(k)=\hbar^2k^2/(2m_i^*)$ is the single nucleon energy expressed with the effective mass of a nucleon $m^*_i$, 
and $\mu_{0i}$ is chosen so as to satisfy nucleon number conservation.    
On the other hand, to obtain the internal energy per nucleon $U_{\mathrm{N}}(n_{\mathrm{p}}, n_{\mathrm{n}},T)$, 
we replace $n_{0i}(k)$ in the calculation of $E(n_{\mathrm{p}}, n_{\mathrm{n}})$ at zero temperature by $n_{\mathrm{T}i}(k)$ for finite temperature. 
Here, the three-body energy $E_3(n_{\mathrm{p}},n_{\mathrm{n}})$ is assumed to be independent of temperature; the validity of this assumption is discussed in Ref.~\cite{K1}.  
Then, we calculate $F_{\mathrm{N}}(n_{\mathrm{p}},n_{\mathrm{n}}, T)$ by minimizing $U_{\mathrm{N}}(n_{\mathrm{p}},n_{\mathrm{n}},T)-TS_{\mathrm{N}}(n_{\mathrm{p}},n_{\mathrm{n}},T)$ with respect to $m^*_i$.  
This means that, in this minimization, the frozen correlation approximation is adopted, 
i.e., the correlation functions $f^{\mu}_{\mathrm{C}ts}(r)$, $f^{\mu}_{\mathrm{T}t}(r)$, $f^{\mu}_{\mathrm{SO}t}(r)$ are fixed to those at zero temperature.  
The validity of this approximation is shown in Ref.~\cite{T1}.  
We note that the validity of the prescription by Schmidt and Pandharipande was investigated by Mukherjee and Pandharipande~\cite{MP}. 
Following this prescription, Mukherjee calculated free energies of pure neutron matter and symmetric nuclear matter~\cite{AM} as extensions of the EOS by APR at zero temperature.  
The free energies per nucleon of pure neutron matter and symmetric nuclear matter derived in this study are in good agreement with these FHNC results.  
Furthermore, the internal energy per nucleon and the entropy per nucleon derived from $F_{\mathrm{N}}(n_{\mathrm{p}},n_{\mathrm{n}},T)$ through thermodynamic relations 
are found to be in excellent agreement with $U_{\mathrm{N}}(n_{\mathrm{p}},n_{\mathrm{n}},T)$ and $S_{\mathrm{N}}(n_{\mathrm{p}},n_{\mathrm{n}},T)$ calculated above.  
This fact implies that the present variational calculation is self-consistent.  

\subsection{Modification of the EOS of uniform nuclear matter at low densities}
To develop a complete nuclear EOS table for supernova simulations, in this paper, we construct an EOS of non-uniform nuclear matter with the Thomas-Fermi approximation.  
In this calculation, we employ the free energy per nucleon $F_{\mathrm{N}}(n_{\mathrm{p}},n_{\mathrm{n}},T)$ of uniform nuclear matter calculated above.  
Before doing so, however, a modification of $F_{\mathrm{N}}(n_{\mathrm{p}},n_{\mathrm{n}},T)$ is necessary to overcome the following difficulty.  

\begin{figure}[t]
  \centering
  \includegraphics[width=8.0cm]{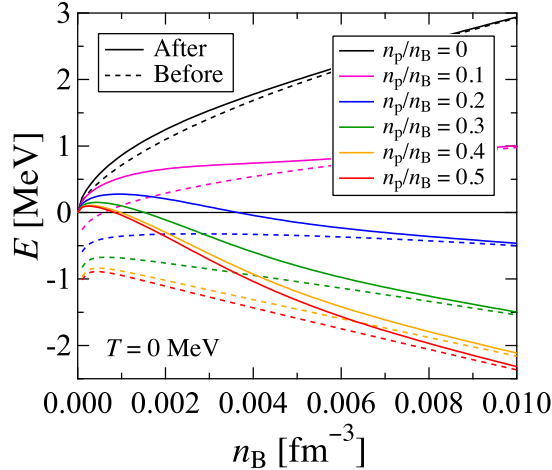}
  \caption{
  (Color online) Energy per nucleon of uniform nuclear matter at various values of proton fraction as a function of nucleon number density $n_{\mathrm{B}}$. 
  Dashed and solid lines represent the results with $r_{\mathrm{h}}$ before and after the modification, respectively. 
  The lines correspond, from top to bottom, to the cases $n_{\mathrm{p}}/n_{\mathrm{B}} = 0$, 0.1, 0.2, 0.3, 0.4, and 0.5. 
  }
\label{fig:Eheal1}
\end{figure}

Figure \ref{fig:Eheal1} shows the energy per nucleon of cold uniform nuclear matter, $E(n_{\mathrm{p}},n_{\mathrm{n}})$, at low densities.  
As $n_{\mathrm{B}}$ approaches zero, the energy per nucleon of neutron matter $E(0,n_{\mathrm{B}})$ approaches zero (the black dashed line).  
In the case of nuclear matter with $n_{\mathrm{p}} > 0$, on the other hand, $E(n_{\mathrm{p}},n_{\mathrm{n}})$ approaches a negative finite value (colored dashed lines).  
In particular, in the case of symmetric nuclear matter, $E(n_{\mathrm{B}}/2,n_{\mathrm{B}}/2)$ approaches $-1.1$ MeV, which corresponds to a half of the deuteron binding energy.  

\begin{figure}[t]
  \centering
  \includegraphics[width=8.0cm]{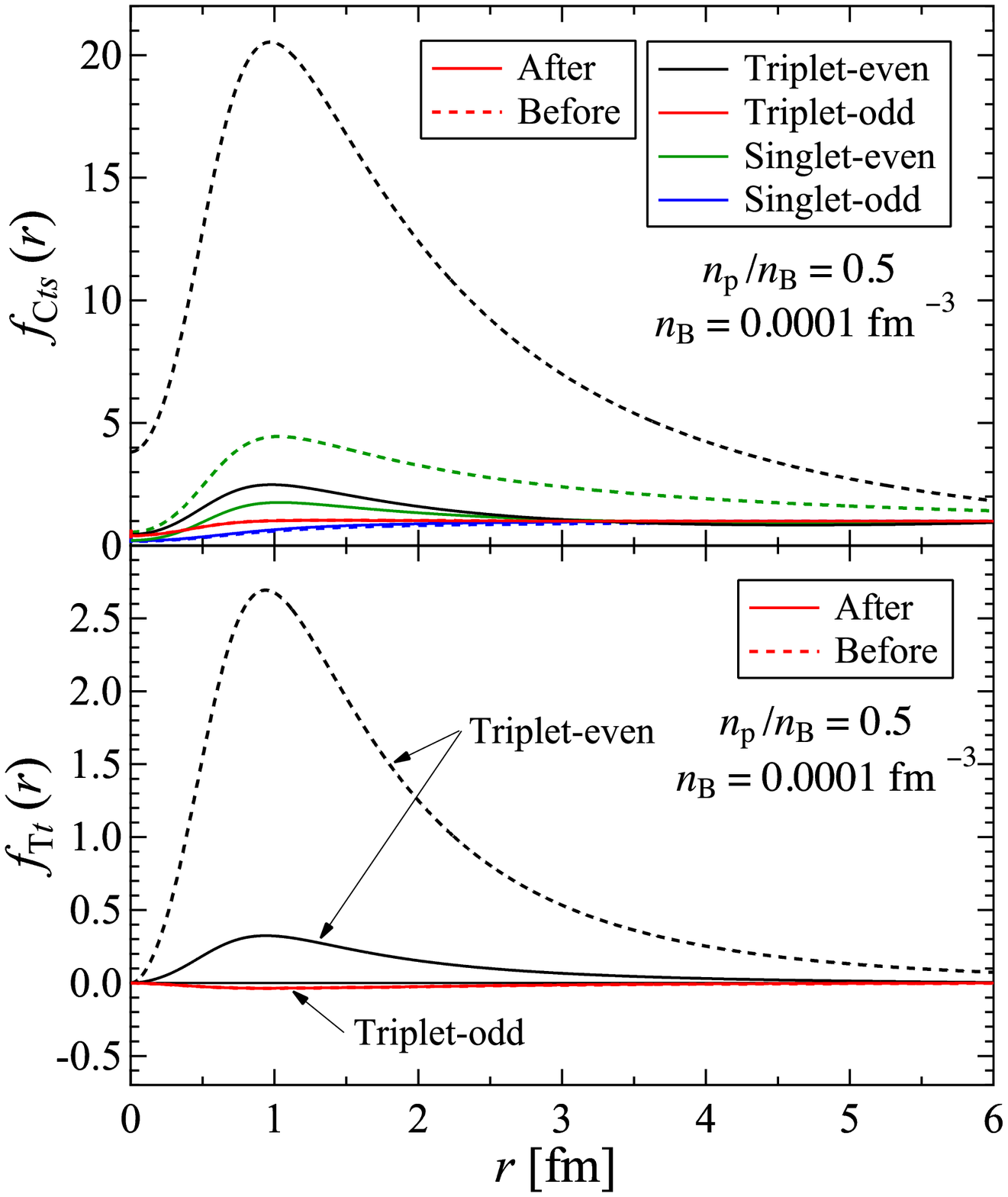}
  \caption{(Color online) Spin-isospin-dependent central (top panel) and tensor (bottom panel) correlation functions for symmetric nuclear matter 
  as functions of the distance between two nucleons $r$ at $n_{\mathrm{B}} = 0.0001$ fm$^{-3}$. 
  Dashed and solid lines represent the results with $r_{\mathrm{h}}$ before and after the modification, respectively. 
  In the top panel, the lines for $r =1$ fm correspond, from top to bottom, to triplet-even [$f^{\mu}_{\mathrm{C}01}(r)$], singlet-even [$f^{\mu}_{\mathrm{C}10}(r)$], 
  triplet-odd [$f^{\mu}_{\mathrm{C}11}(r)$], and singlet-odd [$f^{\mu}_{\mathrm{C}00}(r)$] channels.}
\label{fig:correlation}
\end{figure}

Figure \ref{fig:correlation} shows the correlation functions $f^{\mu}_{\mathrm{C}ts}(r)$ and $f^{\mu}_{\mathrm{T}t}(r)$ for symmetric nuclear matter at $n_{\mathrm{B}}=0.000	1$ fm$^{-3}$.  
It is seen that the amplitudes of $f^{\mu}_{\mathrm{C}01}(r)$ and $f^{\mu}_{\mathrm{T}0}(r)$ are extremely large.  
These results imply that two nucleons tend to form a cluster in the relative triplet-even channel with a finite tensor correlation, i.e., a deuteron.  
This tendency is reasonable because, in principle, nuclear matter, which is a mixture of protons and neutrons, is unstable against formation of nucleon clusters at subnuclear densities. 
In particular, since we adopt the two-body cluster approximation in the variational many-body calculation, the correlation with respect to two-nucleon clustering in nuclear matter is treated rigorously.  
In uniform nuclear matter at a density above the saturation density, 
the correlation between nucleons in uniform nuclear matter vanishes as the distance between two nucleons $r$ approaches the healing distance $r_{\mathrm{h}}=a_{\mathrm{h}}r_0$.  
As the density decreases, $r_{\mathrm{h}}$ becomes larger, and the constraints on the correlation functions are weakened.  
Then, at the low-density limit with $r_{\mathrm{h}}\rightarrow\infty$, the correlation functions in the case of symmetric nuclear matter reduce to the relative wave function of a free deuteron.  
Correspondingly, $-E(n_{\mathrm{B}}/2,n_{\mathrm{B}}/2)$ for symmetric nuclear matter approaches the empirical deuteron binding energy per nucleon 
because we employ the modern realistic two-body nuclear potential AV18 which reproduces deuteron properties.  
The free energy per nucleon $F_{\mathrm{N}}(n_{\mathrm{p}},n_{\mathrm{n}},T)$ at finite temperature also involves the effect of the deuteron cluster formation.  

Although the formation of deuteron clusters is a natural consequence of the present variational method, in constructing an EOS for supernova simulations, 
we treat cluster formation in the subsequent Thomas-Fermi calculations, where we utilize the free energy density of (hypothetical) uniform nuclear matter.  
Thus, for our purpose, it is necessary to prepare the free energy per nucleon $F_{\mathrm{N}}(n_{\mathrm{p}},n_{\mathrm{n}},T)$ without the effect of the deuteron cluster formation.  

To avoid deuteron cluster formation, we modify the healing distance condition as follows: 
\begin{equation}
r_{\mathrm{h}}=\frac{a_{\mathrm{h}}r_0}{\left[1+\left(a_{\mathrm{h}}r_0/b_{\mathrm{h}}\right)^{c_{\mathrm{h}}}\right]^{1/{c_{\mathrm{h}}}}}.  
\label{eq:heal}
\end{equation} 
The healing distance $r_{\mathrm{h}}$ defined above behaves approximately as $a_{\mathrm{h}}r_0$ at relatively high densities and approaches $b_{\mathrm{h}}$ at low densities: 
The latter behavior of $r_{\mathrm{h}}$ significantly suppresses formation of deuteron clusters. 
The parameters $b_{\mathrm{h}}$ and $c_{\mathrm{h}}$ are chosen so that the modified $E(n_{\mathrm{p}},n_{\mathrm{n}})$ is still in good agreement with the results obtained by APR 
and the Thomas-Fermi calculations for isolated atomic nuclei with the modified $E(n_{\mathrm{p}}, n_{\mathrm{n}})$ still reproduce the gross features of their empirical masses and radii.  
The values of the parameters thus obtained are $b_{\mathrm{h}}=$ 6.38 fm and $c_{\mathrm{h}}=10$.  

\begin{figure}[t]
  \centering
  \includegraphics[width=8.0cm]{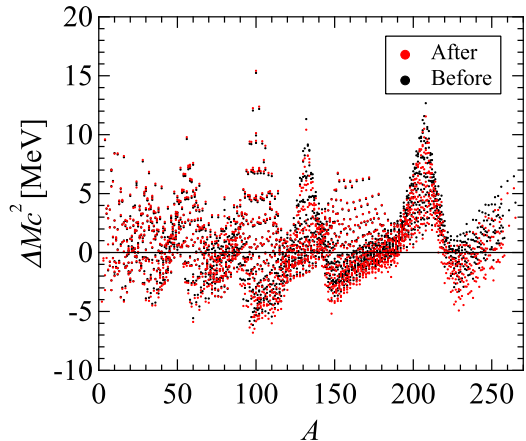}
  \caption{(Color online) Deviations of masses of the mass-measured nuclei obtained by the Thomas-Fermi calculation from the experimental data~\cite{Mexp}.
The black dots represent the results with $r_{\mathrm{h}}$ before the modification, which is taken from Ref.~\cite{K2}.  
The red dots are for the modified $r_{\mathrm{h}}$.}
\label{fig:Mdif}
\end{figure}

The energy per nucleon of uniform nuclear matter at zero temperature with the modified healing distance is shown as solid lines in Fig.~\ref{fig:Eheal1}. 
It is seen that the modified $E(n_{\mathrm{p}}, n_{\mathrm{n}})$ approaches zero as $n_{\mathrm{B}}\rightarrow0$ for both $n_{\mathrm{p}} = 0$ and $n_{\mathrm{p}} > 0$.  
Figure \ref{fig:Mdif} shows deviation of masses of the mass-measured nuclei obtained with the Thomas-Fermi calculation from their experimental mass values~\cite{Mexp}.  
It is seen that the deviations of the modified masses from the experimental data are still sufficiently small; these deviations are mainly caused by the lack of the shell effect in our calculations.  
The RMS deviation of the calculated masses from the experimental data is 2.99 MeV for 2219 nuclides with $Z$ and $N \ge 2$ (2.98 MeV for 2149 nuclides with $Z$ and $N \ge 8$). 
This result is better than that with the unmodified EOS, i.e., 3.09 MeV for 2219 nuclides with $Z$ and $N \ge 2$ (3.08 MeV for 2149 nuclides with $Z$ and $N \ge 8$)~\cite{K2}. 

\begin{figure}[t]
  \centering
  \includegraphics[width=8.0cm]{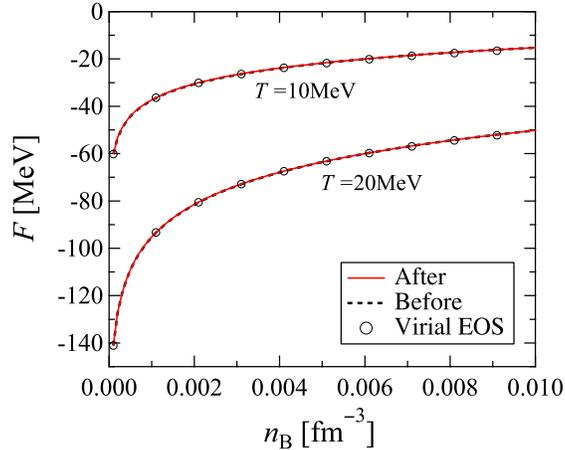}
  \caption{(Color online) Free energy per neutron of uniform neutron matter as a function of nucleon number density $n_{\mathrm{B}}$ at $T$ = 10 and 20 MeV. 
Dashed and solid lines represent the results with $r_{\mathrm{h}}$ before and after the modification, respectively. Open circles show the results of the virial expansion~\cite{virialEOS}.}
\label{fig:virialEOS}
\end{figure}

\begin{figure}[t]
  \centering
  \includegraphics[width=8.0cm]{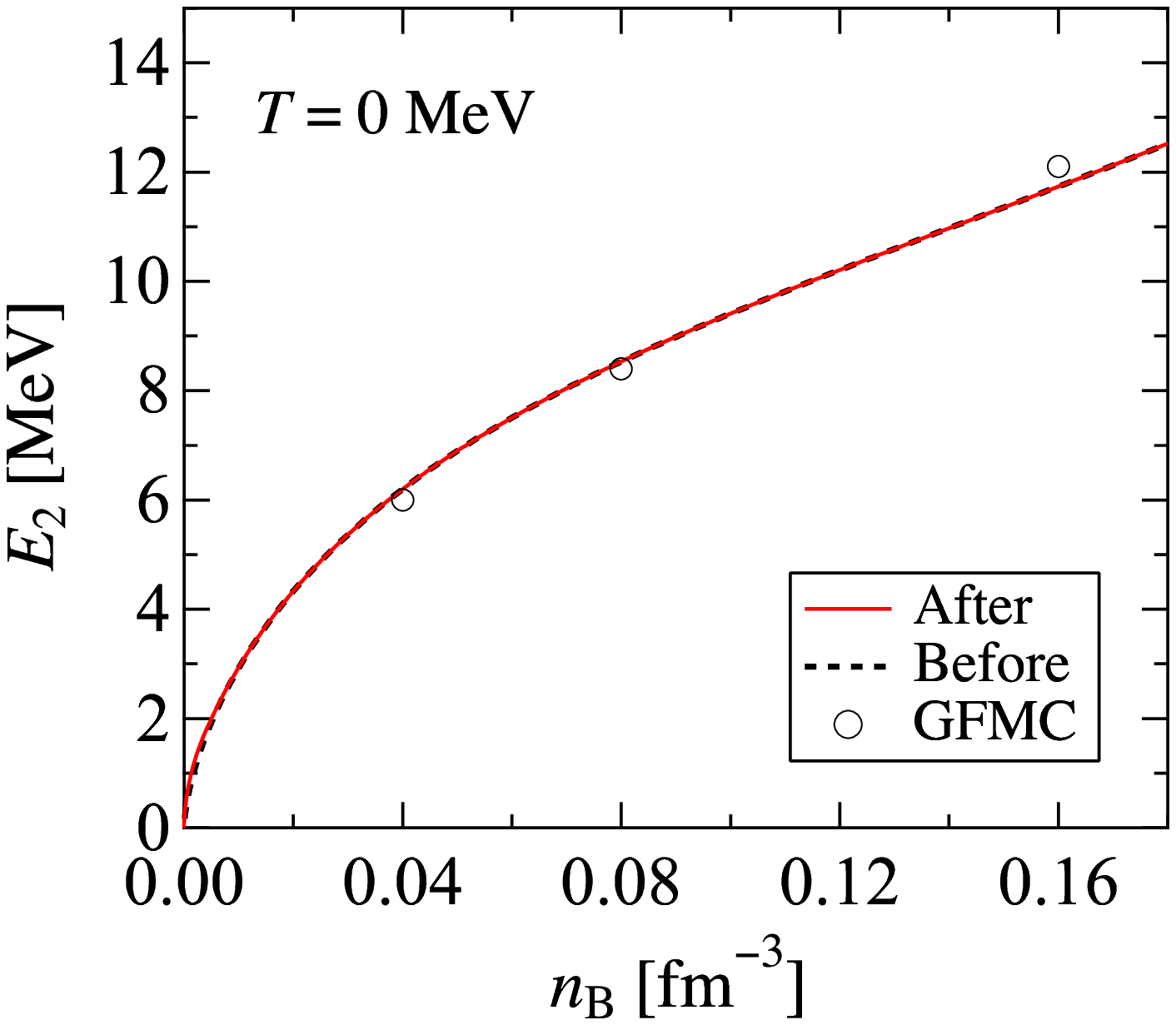}
  \caption{(Color online) Two-body energy per neutron of uniform neutron matter at zero temperature as a function of nucleon number density $n_{\mathrm{B}}$. 
Dashed and solid lines represent the results with $r_{\mathrm{h}}$ before and after the modification, respectively. Open circles show the results of the GFMC calculation based on the AV8$'$ potential~\cite{GFMC}.}
\label{fig:GFMC}
\end{figure}

Here we note that the modification of $r_{\mathrm{h}}$ also affects the free energy per neutron of pure neutron matter $F_{\mathrm{N}}(n_{\mathrm{p}}=0, n_{\mathrm{n}}=n_{\mathrm{B}},T)$ at low densities.  
We maintain the reliability of the EOS for neutron matter as follows. 
First, we refer to the free energies of pure neutron matter calculated with the virial expansion in Ref.~\cite{virialEOS}. 
It is constructed model-independently based on the nucleon-nucleon scattering phase shifts, and thus it is reliable for the thermodynamic states in which the virial expansion is appropriate. 
As shown in Fig.~\ref{fig:virialEOS}, our free energies per neutron of neutron matter are in excellent agreement with those in the virial EOS. 
In addition, at finite temperatures, the differences between the modified and unmodified $F_{\mathrm{N}}(n_{\mathrm{p}},n_{\mathrm{n}},T)$ are so small 
that they are hardly distinguishable from each other in Fig.~\ref{fig:virialEOS}. 
Next, we refer to the energies per neutron of neutron matter at zero temperature calculated with the Green's Function Monte Carlo (GFMC) method~\cite{GFMC} based on the AV8$'$ potential~\cite{AV8}.  
The GFMC method is one of the most reliable techniques to solve a many-body Schr\"odinger equation.  
Here, we note that the AV8$'$ potential used in that study is based on the AV18 potential but composed only of two-body central, tensor, and spin-orbit components of the nuclear force; 
the components that include the second-order relative orbital angular momentum between two nucleons are omitted.  
In fact, the AV8$'$ potential reproduces only the low-angular-momentum isoscalar components of the AV18 potential.  
Furthermore, the three-nucleon force is not taken into account in the GFMC calculation. 
Therefore, the result of the GFMC calculation is more reliable in the low density region than in the high density region, and we refer to them 
so that the two-body energy $E_2(n_{\mathrm{p}}=0, n_{\mathrm{n}}=n_{\mathrm{B}})$ of neutron matter is close to the result of the GFMC calculation at low densities, as shown in Fig.~\ref{fig:GFMC}. 
As expected, it can also be seen that the effect of the modification of $r_{\mathrm{h}}$ on $E_2(n_{\mathrm{p}}=0, n_{\mathrm{n}}=n_{\mathrm{B}})$ is small at relatively high densities. 
With the modified healing distance $r_{\mathrm{h}}$, we have prepared the following free energy density of uniform nuclear matter at zero and finite temperatures 
\begin{equation}
f_{\mathrm{N}}(n_{\mathrm{p}}, n_{\mathrm{n}},T)=(n_{\mathrm{p}}+n_{\mathrm{n}})F_{\mathrm{N}}(n_{\mathrm{p}}, n_{\mathrm{n}},T).  
\label{Funi0}
\end{equation}

We note that the deuteron clustering discussed above does not describe the realistic deuteron mixing in hot dilute nuclear matter as treated in recent other supernova EOS tables \cite{TFhot3, sumi, Hempel2, Furusawa} for the following reasons:  
i) Since we employ the frozen correlation approximation in our variational calculation, the deuteron-type strong correlations at zero temperature survive even at finite temperatures, while most deuterons would be destroyed in realistic hot supernova matter.  
ii) In the case of light nuclei such as deuterons, the energy contribution from the center-of-mass motion would be important.  
However, in our variational method, the center-of-mass motion of deuteron clusters is not taken into account.  
Namely, extensions of our theory are necessary to treat deuteron mixing appropriately in the present EOS, which is beyond our scope. 

\subsection{Mixing of alpha particles and thermodynamic quantities of uniform nuclear matter}
In this subsection, we consider the inclusion of alpha particles which appear at various densities and temperatures in uniform nuclear matter, following the procedure adopted in the Shen EOS. 
For a given set of $n_{\mathrm{p}}$, $n_{\mathrm{n}}$, and the number density of alpha particles $n_{\alpha}$, 
the free energy density of uniform matter $f(n_{\mathrm{p}}, n_{\mathrm{n}}, n_{\alpha},T)$ is expressed as the sum of the contributions from nucleons and alpha particles.  
The contribution from the nucleons $f_{\mathrm{N}}(n_{\mathrm{p}}, n_{\mathrm{n}},T)$ was obtained in the previous subsections, 
while the alpha particle is treated simply as a classical particle with a fixed volume. 
Then, the free energy density of alpha particles is given by
\begin{equation}
f_{\alpha}(n_{\alpha},T) = -T n_{\alpha} \left[ \ln \frac{8n_{\mathrm{Q}}}{n_{\alpha}} +1 \right] - n_{\alpha}B_{\alpha},
\label{Falpha}
\end{equation}
with the abbreviation $n_{\mathrm{Q}}=[m_{\mathrm{n}}T/(2\pi\hbar^2)]^{3/2}$ and the alpha particle binding energy $B_{\alpha}=28.3$~MeV. 
It is noted that we adopt the neutron mass $m_{\mathrm{n}}$ as the nucleon mass, 
which is consistent with the choice of the nucleon mass in the kinetic energy of nucleons in nuclear matter, as mentioned above.  

The excluded volume effect of alpha particles is also taken into account.  
With use of a volume fraction of alpha particles $u=n_{\alpha}v_{\alpha}$ with the effective volume of an alpha particle $v_{\alpha}=24$~fm$^3$, 
the free energy density is obtained by 
\begin{equation}
f(n_{\mathrm{p}}, n_{\mathrm{n}}, n_{\alpha},T) = (1-u)\left[f_{\mathrm{N}}(\tilde{n}_{\mathrm{p}}, \tilde{n}_{\mathrm{n}},T) + f_{\alpha}(\tilde{n}_{\alpha},T)\right], 
\label{Fbulk2}
\end{equation}
where the effective number densities of protons, neutrons, and alpha particles are designated by $\tilde{n}_i=n_i/(1-u)$ with $i= \mathrm{p}$, $\mathrm{n}$, and $\alpha$, respectively. 

To complete the supernova EOS table, we compute the free energy density $f(n_{\mathrm{p}}, n_{\mathrm{n}}, n_{\alpha},T)$
at a given set of temperature $T$, total baryon number density $n_{\mathrm{B}}$, and proton fraction $Y_{\mathrm{p}}$, 
with the latter two expressed by
\begin{equation}
n_{\mathrm{B}} = n_{\mathrm{p}}+n_{\mathrm{n}}+4n_{\alpha}
\label{nBuni}
\end{equation}
and 
\begin{equation}
Y_{\mathrm{p}} = \frac{n_{\mathrm{p}}+2n_{\alpha}}{n_{\mathrm{B}}}.  
\label{Ypuni}
\end{equation}
Then, $f(n_{\mathrm{p}}, n_{\mathrm{n}}, n_{\alpha},T)$ is minimized numerically with respect to $n_{\alpha}$ for a fixed set ($T$, $n_{\mathrm{B}}$, $Y_{\mathrm{p}}$) to determine the optimal $n_{\alpha}$. 
The obtained minimal free energy density is denoted by $f(n_{\mathrm{B}}, Y_{\mathrm{p}},T)$, and the corresponding free energy per baryon including alpha particles is given by
\begin{equation}
F(n_{\mathrm{B}}, Y_{\mathrm{p}},T) = \frac{f(n_{\mathrm{B}}, Y_{\mathrm{p}},T)}{n_{\mathrm{B}}}.  
\label{Funi}
\end{equation}

\section{Thomas-Fermi calculation for non-uniform matter}
In this section, we construct the nuclear EOS for non-uniform matter, 
which appears in the low-temperature and low-density region, based on the nuclear EOS for uniform matter constructed in the last section. 
For this purpose, we adopt the Thomas-Fermi calculation as in Shen et al.~\cite{Shen2, Shen3}, 
which is the extension of the work by Oyamatsu~\cite{Oyak1} to inhomogeneous nuclear matter at finite temperature. 
In this calculation, non-uniform nuclear matter is assumed to be a mixture of free neutrons, free protons, alpha particles, and a single species of heavy nuclei.  
In the Coulomb energy calculation, we also take into account the contribution from a uniform electron gas that coexists with these baryons in supernova matter.  
The goal of our Thomas-Fermi calculation is to determine the thermodynamically favorable state for a system 
with a fixed nucleon number density $n_{\mathrm{B}}$, proton fraction $Y_{\mathrm{p}}$, and temperature $T$.

We assume that a spherical single heavy nucleus is located at the center of a Wigner-Seitz (WS) cell in a body-centered cubic (BCC) lattice. 
For this, we consider the number density distributions in the WS cell for protons $n_\mathrm{p} (r)$, neutrons $n_\mathrm{n} (r)$, and alpha particles $n_\alpha (r)$, 
where $r$ is the distance from the center of the WS cell. 
The number density distributions that minimize the free energy density $F_{\mathrm{cell}}/V_{\mathrm{cell}}$ are favored, 
with $F_{\mathrm{cell}}$ and $V_{\mathrm{cell}}$ being the free energy and the volume of a WS cell, respectively. 
In this model, $F_{\mathrm{cell}}$ is expressed as 
\begin{equation}
F_{\mathrm{cell}}(n_{\mathrm{B}}, Y_{\mathrm{p}},T) = F_{\mathrm{bulk}} + E_{\mathrm{grad}} + E_{\mathrm{C}},
\label{Fcell}
\end{equation}
where $F_{\mathrm{bulk}}$, $E_{\mathrm{grad}}$, and $E_{\mathrm{C}}$ are the bulk free energy, the gradient term, and the Coulomb energy, respectively.
The bulk free energy $F_{\mathrm{bulk}}$ is given by
\begin{equation}
F_{\mathrm{bulk}} = \int_{\mathrm{cell}} d\mbox{\boldmath$r$} f(n_{\mathrm{p}}(r), n_{\mathrm{n}}(r), n_{\alpha}(r),T),  
\label{Fbulk}
\end{equation}
where $f(n_{\mathrm{p}}(r), n_{\mathrm{n}}(r), n_{\alpha}(r),T)$ is the local free energy density given in Eq.~(\ref{Fbulk2}). 
For the gradient term, we adopt the following form:
\begin{equation}
E_{\mathrm{grad}} = F_0 \int_{\mathrm{cell}}  d\mbox{\boldmath$r$}  |\nabla (n_{\mathrm{p}}(r)+n_{\mathrm{n}}(r))|^2,  
\label{Egrad}
\end{equation}
where $F_0$ is chosen to be 68.00 MeV fm$^{-5}$ so that the Thomas-Fermi calculations for isolated atomic nuclei reproduce the gross features of their masses (Fig.~\ref{fig:Mdif}) and radii~\cite{K2}. 
The Coulomb energy is expressed as
\begin{eqnarray}
E_{\mathrm{C}} &=& \frac{e^2}{2} \int_{\mathrm{cell}} d\mbox{\boldmath$r$} \int_{\mathrm{cell}}  d\mbox{\boldmath$r'$} 
\frac{[n_{\mathrm{p}}(r)+2n_{\alpha}(r)-n_{\mathrm{e}}][n_{\mathrm{p}}(r')+2n_{\alpha}(r')-n_{\mathrm{e}}]}
{|\mbox{\boldmath$r$}-\mbox{\boldmath$r'$}|} \nonumber  \\
&+& c_{\mathrm{bcc}}\frac{(Z_{\mathrm{non}}e)^2}{a},
\label{FC}
\end{eqnarray}
where $n_{\mathrm{e}}= Y_{\mathrm{p}}n_{\mathrm{B}}$ is the electron number density of a uniform electron gas. 
Note that, in our Thomas-Fermi calculation, the average nucleon number density is defined by
\begin{equation}
n_{\mathrm{B}} = \frac{1}{V_{\mathrm{cell}}}\int_{\mathrm{cell}} d\mbox{\boldmath$r$} \, \bigl\{ n_{\mathrm{p}}(r)+n_{\mathrm{n}}(r)+4n_{\alpha}(r) \bigr\},
\label{nBnonuni}
\end{equation}
and the average proton fraction is defined by
\begin{equation}
Y_{\mathrm{p}} = \frac{\int_{\mathrm{cell}} d\mbox{\boldmath$r$} \, \bigl\{ n_{\mathrm{p}}(r)+2n_{\alpha}(r) \bigr\}}{\int_{\mathrm{cell}} d\mbox{\boldmath$r$} \, \bigl\{ n_{\mathrm{p}}(r)+n_{\mathrm{n}}(r)+4n_{\alpha}(r) \bigr\}}.
\label{Ypnonuni}
\end{equation}
The last term on the right-hand side of Eq.~(\ref{FC}) is the modification of the Coulomb energy for the BCC lattice. 
In this term, $a$ is the lattice constant defined by $V_{\mathrm{cell}} = a^3$, $Z_{\mathrm{non}}$ is the non-uniform part of the charge number per cell, and $c_{\mathrm{bcc}}$ is chosen to be 0.006562~\cite{Oyak1}.

In order to determine the thermodynamically favorable state, we further assume the functional form of the particle number density distributions in a WS cell. 
It is parameterized as
\begin{eqnarray}
n_i(r) = \left\{ \begin{array}{ll}
(n_i^{\mathrm{in}}-n_i^{\mathrm{out}})[1 - (r/R_i)^{t_i}]^3 + n_i^{\mathrm{out}}   &(0 \le r \le R_i), \\
n_i^{\mathrm{out}}    &(R_i \le r \le R_{\mathrm{cell}}),      \label{eq1}
\end{array}
\right. 
\end{eqnarray}
for protons ($i = \mathrm{p}$) and neutrons ($i = \mathrm{n}$). 
Here $R_{\mathrm{cell}}$ is the radius of the WS cell defined by $V_{\mathrm{cell}} = 4 \pi R_{\mathrm{cell}}^3/3$, and $R_{\mathrm{n}}$ and $R_{\mathrm{p}}$ are the radii of the neutron and proton distributions, respectively, 
in the heavy nucleus situated in the WS cell. 
The distribution of alpha particles is assumed to be 
\begin{eqnarray}
n_{\alpha}(r) = \left\{ \begin{array}{ll}
-n_{\alpha}^{\mathrm{out}}[1-(r/R_{\mathrm{p}})^{t_{\mathrm{p}}}]^3 + n_{\alpha}^{\mathrm{out}}  &(0 \le r \le R_{\mathrm{p}}), \\
n_{\alpha}^{\mathrm{out}}    &(R_{\mathrm{p}} \le r \le R_{\mathrm{cell}}).    \label{eq2}
\end{array}
\right.
\end{eqnarray}
We remark that, using these parameterized density distribution functions, the non-uniform part of the charge number per cell, $Z_{\mathrm{non}}$, is given by
\begin{equation}
Z_{\mathrm{non}} = 4\pi \int_0^{R_{\mathrm{p}}} 
(n_{\mathrm{p}}^{\mathrm{in}}-n_{\mathrm{p}}^{\mathrm{out}}-2n_{\mathrm{\alpha}}^{\mathrm{out}}) 
\left[1 - \Bigg(\frac{r}{R_{\mathrm{p}}} \Bigg)^{t_{\mathrm{p}}} \right]^3 r^2 dr.
\end{equation}
Then, the average free energy density of a WS cell $F_{\mathrm{cell}}/V_{\mathrm{cell}}$ is minimized numerically with respect to eight independent parameters among 
$a$, $n_{\mathrm{n}}^{\mathrm{in}}$, $n_{\mathrm{n}}^{\mathrm{out}}$, $R_{\mathrm{n}}$, $t_{\mathrm{n}}$, $n_{\mathrm{p}}^{\mathrm{in}}$, $n_{\mathrm{p}}^{\mathrm{out}}$, $R_{\mathrm{p}}$, $t_{\mathrm{p}}$, and $n_{\alpha}^{\mathrm{out}}$ for each $n_{\mathrm{B}}$, $Y_{\mathrm{p}}$ and $T$. 
Note that the minimization of $F_{\mathrm{cell}}/V_{\mathrm{cell}}$ with respect to the lattice constant $a$ provides us with the following equation~\cite{Oyak1}:
\begin{equation}
E_{\mathrm{grad}} = E_{\mathrm{C}}.
\label{sizeeqlbr}
\end{equation}
We use this condition to confirm that sufficient numerical minimization is achieved.  

The free energy per baryon for non-uniform matter is calculated from the optimized $F_{\mathrm{cell}}/V_{\mathrm{cell}}$ as
\begin{equation}
F(n_{\mathrm{B}}, Y_{\mathrm{p}},T) = \frac{1}{n_{\mathrm{B}}} \left( \frac{F_{\mathrm{cell}}}{V_{\mathrm{cell}}} \right).
\label{Fnonuni}
\end{equation}
Furthermore we define the mass number $A$ and charge number $Z$ of the heavy nuclei in the Thomas-Fermi calculation of non-uniform matter as
\begin{eqnarray}
A &=& 4\pi \int_0^{R_{\mathrm{A}}} [n_{\mathrm{p}}(r)+n_{\mathrm{n}}(r)] \, r^2dr, \label{defA} \\
Z &=& 4\pi \int_0^{R_{\mathrm{A}}} n_{\mathrm{p}}(r) \, r^2dr, \label{defZ}
\end{eqnarray}
where $R_{\mathrm{A}}$ is regarded as the radius of the heavy nucleus and defined by the maximum of $R_{\mathrm{p}}$ and $R_{\mathrm{n}}$. 
The fraction of heavy nuclei is defined by
\begin{equation}
X_{\mathrm{A}} = \frac{A}{n_{\mathrm{B}}V_{\mathrm{cell}}}.
\label{XA}
\end{equation}
For non-uniform matter at high $T$, high $n_{\mathrm{B}}$, or low $Y_{\mathrm{p}}$, protons and/or neutrons drip out of the heavy nuclei. 
Number densities of these free protons and neutrons correspond to $n_{\mathrm{p}}^{\mathrm{out}}$ and $n_{\mathrm{n}}^{\mathrm{out}}$, respectively, in our Thomas-Fermi model. 
Then, the free proton fraction $X_{\mathrm{p}}$ and free neutron fraction $X_{\mathrm{n}}$ are given by
\begin{equation}
X_{i} = \frac{n_{i}^{\mathrm{out}} V^{\mathrm{out}}}{n_{\mathrm{B}}V_{\mathrm{cell}}},
\label{XiTF}
\end{equation}
with $i = \mathrm{p}, \mathrm{n}$. Here $V^{\mathrm{out}}=V_{\mathrm{cell}}-4 \pi R_{\mathrm{A}}^3/3$ is the volume outside the nucleus. 
The alpha-particle fraction is defined by
\begin{equation}
X_{\alpha} = \frac{4}{n_{\mathrm{B}}V_{\mathrm{cell}}}\int_{\mathrm{cell}} d\mbox{\boldmath$r$} \, n_{\alpha}(r).
\label{XalphaTF}
\end{equation}

By comparing the free energy $F(n_{\mathrm{B}}, Y_{\mathrm{p}},T)$ for non-uniform nuclear matter in Eq.~(\ref{Fnonuni}) with that for uniform nuclear matter in Eq.~(\ref{Funi}), 
we finally choose the most favorable value of the free energy $F(n_{\mathrm{B}}, Y_{\mathrm{p}},T)$ for each $n_{\mathrm{B}}$, $Y_{\mathrm{p}}$ and $T$. 
We then derive the thermodynamic quantities listed in the EOS table from $F(n_{\mathrm{B}}, Y_{\mathrm{p}},T)$. 
In particular, the pressure $P(n_{\mathrm{B}}, Y_{\mathrm{p}},T)$, the entropy per baryon $S(n_{\mathrm{B}}, Y_{\mathrm{p}},T)$, 
and the internal energy per baryon $U(n_{\mathrm{B}}, Y_{\mathrm{p}},T)$ of uniform nuclear matter are given by
\begin{eqnarray}
P(n_{\mathrm{B}}, Y_{\mathrm{p}},T) &=& n_{\mathrm{B}}^2 \left[ \frac{\partial F(n_{\mathrm{B}}, Y_{\mathrm{p}},T)}{\partial n_{\mathrm{B}}} \right]_{Y_{\mathrm{p}},T}, \label{themP} \\
S(n_{\mathrm{B}}, Y_{\mathrm{p}},T) &=& -\left[ \frac{\partial F(n_{\mathrm{B}}, Y_{\mathrm{p}},T)}{\partial T} \right]_{n_{\mathrm{B}}, Y_{\mathrm{p}}}, \label{themS} \\
U(n_{\mathrm{B}}, Y_{\mathrm{p}},T) &=& F(n_{\mathrm{B}}, Y_{\mathrm{p}},T) + TS(n_{\mathrm{B}}, Y_{\mathrm{p}},T),  \label{themU}
\end{eqnarray}
respectively.  
Here, the internal energy per baryon given in the EOS table, $E_{\mathrm{int}}(n_{\mathrm{B}}, Y_{\mathrm{p}},T)$, is measured relative to the atomic mass unit $m_{\mathrm{u}}$ times the speed of light squared $c^2$ as in the Shen EOS table:
\begin{equation}
E_{\mathrm{int}}(n_{\mathrm{B}}, Y_{\mathrm{p}},T) = U(n_{\mathrm{B}}, Y_{\mathrm{p}},T) + Y_{\mathrm{p}}m_{\mathrm{p}}c^2+ (1-Y_{\mathrm{p}})m_{\mathrm{n}}c^2 -m_{\mathrm{u}}c^2.
\label{themEint}
\end{equation}
The chemical potentials of protons $\mu_{\mathrm{p}}(n_{\mathrm{B}}, Y_{\mathrm{p}},T)$ and neutrons $\mu_{\mathrm{n}}(n_{\mathrm{B}}, Y_{\mathrm{p}},T)$ are derived from 
the free energy density $f(n_{\mathrm{B}}, Y_{\mathrm{p}},T) = n_{\mathrm{B}}F(n_{\mathrm{B}}, Y_{\mathrm{p}},T)$ by 
\begin{eqnarray}
\mu_{\mathrm{p}}(n_{\mathrm{B}}, Y_{\mathrm{p}},T) &=& \left[ \frac{\partial f(n_{\mathrm{B}}, Y_{\mathrm{p}},T)}{\partial \bar n_{\mathrm{p}}} \right]_{\bar n_{\mathrm{n}},T}, \label{themMUp} \\
\mu_{\mathrm{n}}(n_{\mathrm{B}}, Y_{\mathrm{p}},T) &=& \left[ \frac{\partial f(n_{\mathrm{B}}, Y_{\mathrm{p}},T)}{\partial \bar n_{\mathrm{n}}} \right]_{\bar n_{\mathrm{p}},T}, \label{themMUn}
\end{eqnarray}
where $\bar n_{\mathrm{p}}=Y_{\mathrm{p}}n_{\mathrm{B}}$ and $\bar n_{\mathrm{n}}=(1-Y_{\mathrm{p}})n_{\mathrm{B}}$ are the average number densities of protons and neutrons, respectively. 

\section{Properties of the EOS for non-uniform matter}
In this section, we study the phase transition between non-uniform matter and uniform matter. 
The results are compared with the work by Shen et al.~\cite{Shen2, Shen3}. 

\begin{figure}[t]
  \centering
  \includegraphics[width=7.0cm]{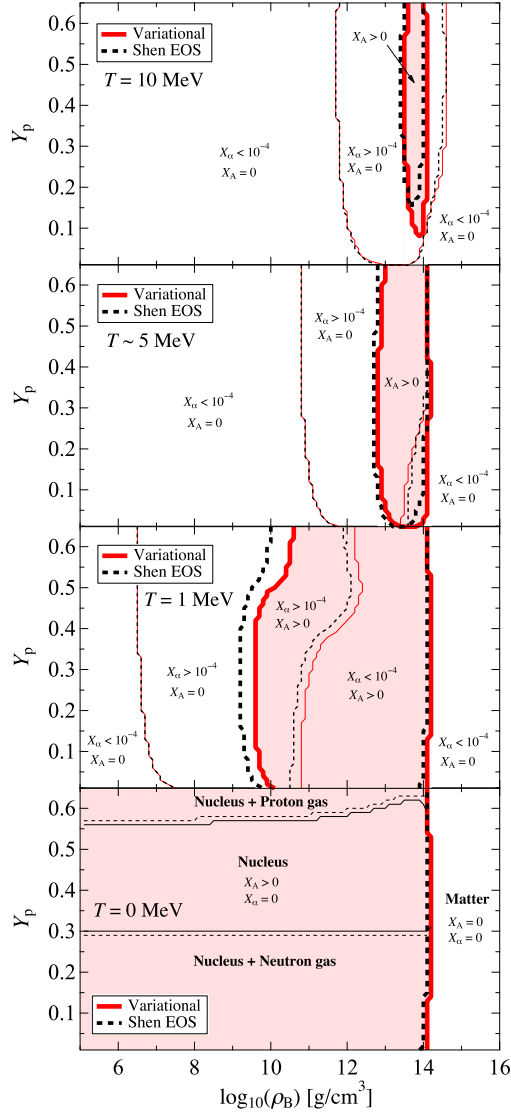}
  \caption{(Color online) Phase diagram of nuclear matter at $T$ = 0, 1, 5, and 10 MeV (from bottom to top panels) in the $\rho_{\mathrm{B}}$-$Y_{\mathrm{p}}$ plane.
The shaded region represents the non-uniform phase in which heavy nuclei are formed. 
The thin lines at finite temperatures correspond to the alpha-particle fraction $X_{\alpha} = 10^{-4}$, representing the boundaries of alpha-mixed regions. 
The neutron and proton drip lines are also shown at zero temperature by thin lines. 
Solid and dashed lines are for our EOS and for the Shen EOS, respectively.}
\label{fig:rhoBYp}
\end{figure}
\clearpage

The resultant phase diagrams in the $\rho_{\mathrm{B}}$-$Y_{\mathrm{p}}$ plane are shown in Fig.~\ref{fig:rhoBYp}. 
Here, the baryon mass density is defined by $\rho_{\mathrm{B}}=m_{\mathrm{u}}n_{\mathrm{B}}$ with $m_{\mathrm{u}}$ being the atomic mass unit, following the definition by Shen et al.~\cite{Shen2, Shen3}. 

The top three panels in Fig.~\ref{fig:rhoBYp} show the phase diagrams at finite temperatures $T$ = 1, $10^{0.68} \sim 5$, and 10 MeV.
It is seen in this figure that the system at a finite temperature is a homogeneous nucleon gas at low densities; 
as the density increases, the non-uniform phase appears, and then nuclear matter becomes uniform at high densities. 
In analogy with the liquid-vapor transition of water, the non-uniform phase at medium densities is regarded as the coexistence phase, 
while homogeneous phases at high and low densities correspond to liquid and vapor, respectively. 
At $T$=1 MeV, the transition densities from the uniform to non-uniform phase in our EOS are higher than those of the Shen EOS. 
This tendency is consistent with the fact that 
the saturation density of symmetric nuclear matter for our EOS ($n_0$ = 0.16 fm$^{-3}$) is higher than that for the Shen EOS ($n_0$ = 0.145 fm$^{-3}$) at zero temperature. 
It is also seen in Fig.~\ref{fig:rhoBYp} that, in the low-$Y_{\mathrm{p}}$ region, the transition densities from non-uniform to uniform phase in our EOS are higher than those of the Shen EOS. 
This is because the density derivative coefficient of the symmetry energy of our EOS ($L=35$~MeV) is smaller than that of the Shen EOS ($L=111$~MeV) or, 
in other words, the symmetry energy at subnuclear densities is larger in our EOS. 
The phase separation is induced by the proton clustering, which is driven in part by the symmetry energy at subnuclear densities~\cite{Oyak2}.  
As a result, our EOS is less stable against proton clustering and has higher transition density in the low-$Y_{\mathrm{p}}$ region. 
In addition, the thin lines at finite temperatures correspond to the alpha-particle fraction $X_{\alpha} = 10^{-4}$, representing the boundaries of the alpha-mixed region. 
It is seen that the boundaries in our EOS are very close to those in the Shen EOS. 

\begin{figure}[t]
  \centering
  \includegraphics[width=7.0cm]{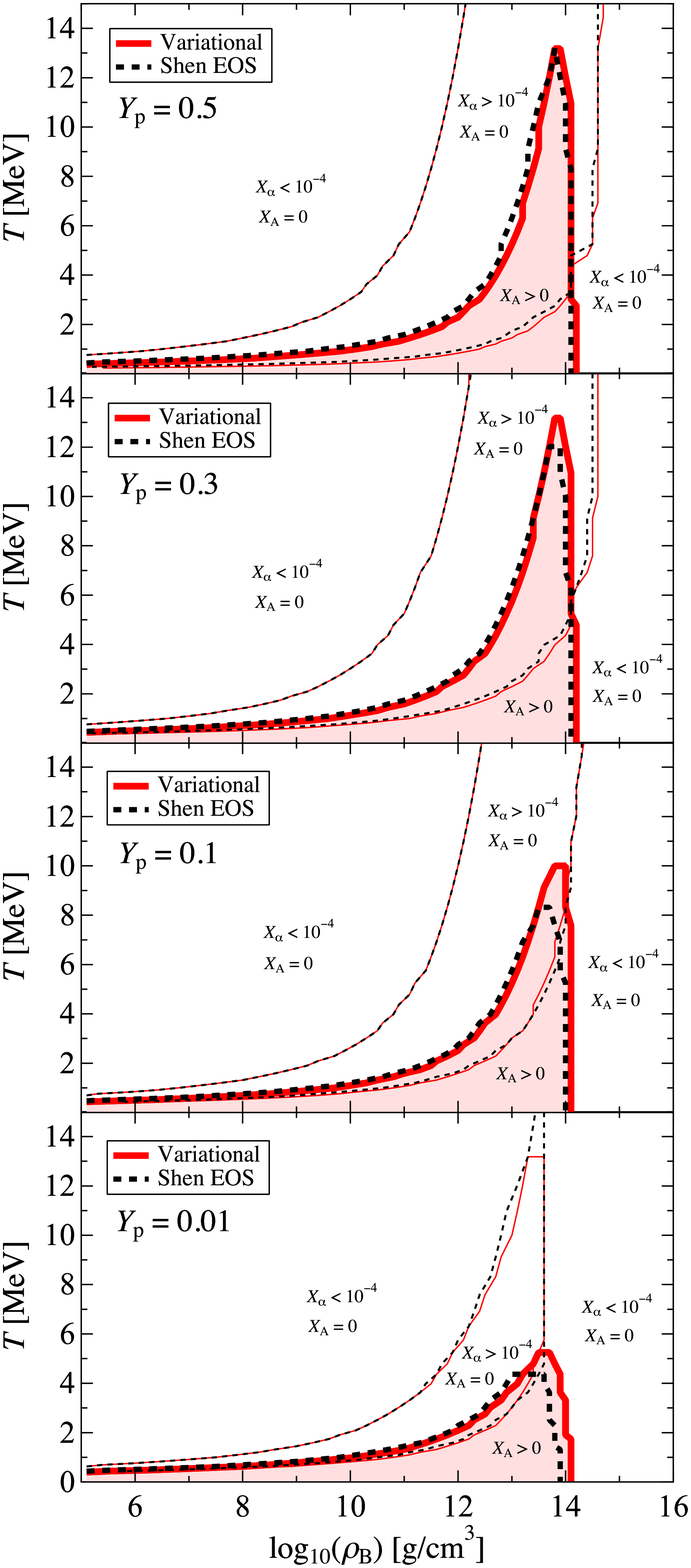}
  \caption{(Color online) Phase diagram of nuclear matter at $Y_{\mathrm{p}}=0.01$, 0.1, 0.3, and 0.5 (from bottom to top panels) in the $\rho_{\mathrm{B}}$-$T$ plane.
The shaded region represents the non-uniform phase in which heavy nuclei are formed. 
The thin lines correspond to the alpha-particle fraction $X_{\alpha} = 10^{-4}$, representing the boundaries of alpha-mixed region. 
The solid and dashed lines are for our EOS and for the Shen EOS, respectively.}
\label{fig:rhoBT}
\end{figure}
\clearpage

Figure~\ref{fig:rhoBT} shows phase diagrams in the $\rho_{\mathrm{B}}$-$T$ plane. 
It is seen that, as the temperature increases, heavy nuclei disintegrate into alpha particles and nucleons, and finally the system becomes a homogeneous nucleon gas. 
The phase of homogeneous nucleon gas is connected continuously to the phase of uniform nuclear matter in the high-temperature region. 
It is also seen that the non-uniform region becomes smaller as $Y_{\mathrm{p}}$ decreases, and correspondingly, the critical temperature goes down. 
For symmetric nuclear matter, the critical temperature for our EOS is close to that for the Shen EOS.
As $Y_{\mathrm{p}}$ decreases, the critical temperature in our EOS decreases more moderately than in the Shen EOS. 
This tendency is understood from the following consideration: 
The pressure of nuclear matter $P(n_{\mathrm{B}},Y_{\mathrm{p}},T)$ is approximated by the sum of the pressure at zero temperature $P_0(n_{\mathrm{B}},Y_{\mathrm{p}})$ 
and thermal pressure as $P(n_{\mathrm{B}},Y_{\mathrm{p}},T) \sim P_0(n_{\mathrm{B}},Y_{\mathrm{p}})+n_{\mathrm{B}}T$, in which $P_0(n_{\mathrm{B}},Y_{\mathrm{p}})$ 
is evaluated from the energy per baryon at zero temperature $E(n_{\mathrm{B}},Y_{\mathrm{p}})$ 
as $P_0(n_{\mathrm{B}},Y_{\mathrm{p}}) = n_{\mathrm{B}}^2(\partial E(n_{\mathrm{B}},Y_{\mathrm{p}})/\partial n_{\mathrm{B}})_{Y_{\mathrm{p}}}$. 
It is known that $E(n_{\mathrm{B}},Y_{\mathrm{p}})$ around $n_{\mathrm{B}}=n_0$ is expressed approximately as
\begin{equation}
E(n_{\mathrm{B}},Y_{\mathrm{p}}) \sim E_0 + \frac{K}{18n_0^2}(n_{\mathrm{B}}-n_0)^2+\left[ E_{\mathrm{sym}} + \frac{L}{3n_0}(n_{\mathrm{B}}-n_0) \right](1-2Y_{\mathrm{p}})^2,
\label{wTaylor}
\end{equation}
from which we get
\begin{equation}
P(n_{\mathrm{B}},Y_{\mathrm{p}},T) \sim \frac{K}{9n_0^2}(n_{\mathrm{B}}-n_0)n_{\mathrm{B}}^2 + \frac{L}{3n_0}n_{\mathrm{B}}^2(1-2Y_{\mathrm{p}})^2 + n_{\mathrm{B}}T.
\label{Papprox}
\end{equation}
Furthermore, the partial derivative of the pressure with respect to $n_{\mathrm{B}}$ is given by
\begin{equation}
\left.\frac{\partial P(n_{\mathrm{B}},Y_{\mathrm{p}},T)}{\partial n_{\mathrm{B}}}\right |_{Y_{\mathrm{p}},T} \sim \frac{K}{9n_0^2}(3n_{\mathrm{B}}-2n_0)n_{\mathrm{B}} + \frac{2L}{3n_0}n_{\mathrm{B}}(1-2Y_{\mathrm{p}})^2 + T.
\label{dPdn}
\end{equation}
For a system with fixed $Y_{\mathrm{p}}$ and $T$, the non-uniform phase appears if $\partial P(n_{\mathrm{B}},Y_{\mathrm{p}},T) /\partial n_{\mathrm{B}}$ is negative at some $n_{\mathrm{B}}$. 
As seen in Eq.~(\ref{dPdn}), $\partial P(n_{\mathrm{B}},Y_{\mathrm{p}},T) /\partial n_{\mathrm{B}}$ depends on $Y_{\mathrm{p}}$ through $L$, i.e., $\partial P(n_{\mathrm{B}},Y_{\mathrm{p}},T) /\partial n_{\mathrm{B}}$ varies more largely with larger $L$.  
This tendency also holds for nuclear matter at subnuclear densities where non-uniform phase appears.  
Since $L$ in our EOS is smaller than that in the Shen EOS, 
the $Y_{\mathrm{p}}$ dependences of $\partial P(n_{\mathrm{B}},Y_{\mathrm{p}},T) /\partial n_{\mathrm{B}}$ and the related critical temperature are more moderate for our EOS. 

\begin{center}
\begin{figure}[t]
\begin{tabular}{cc}
\begin{minipage}[t]{0.48\hsize} 
\centering
\includegraphics[width=6.5cm]{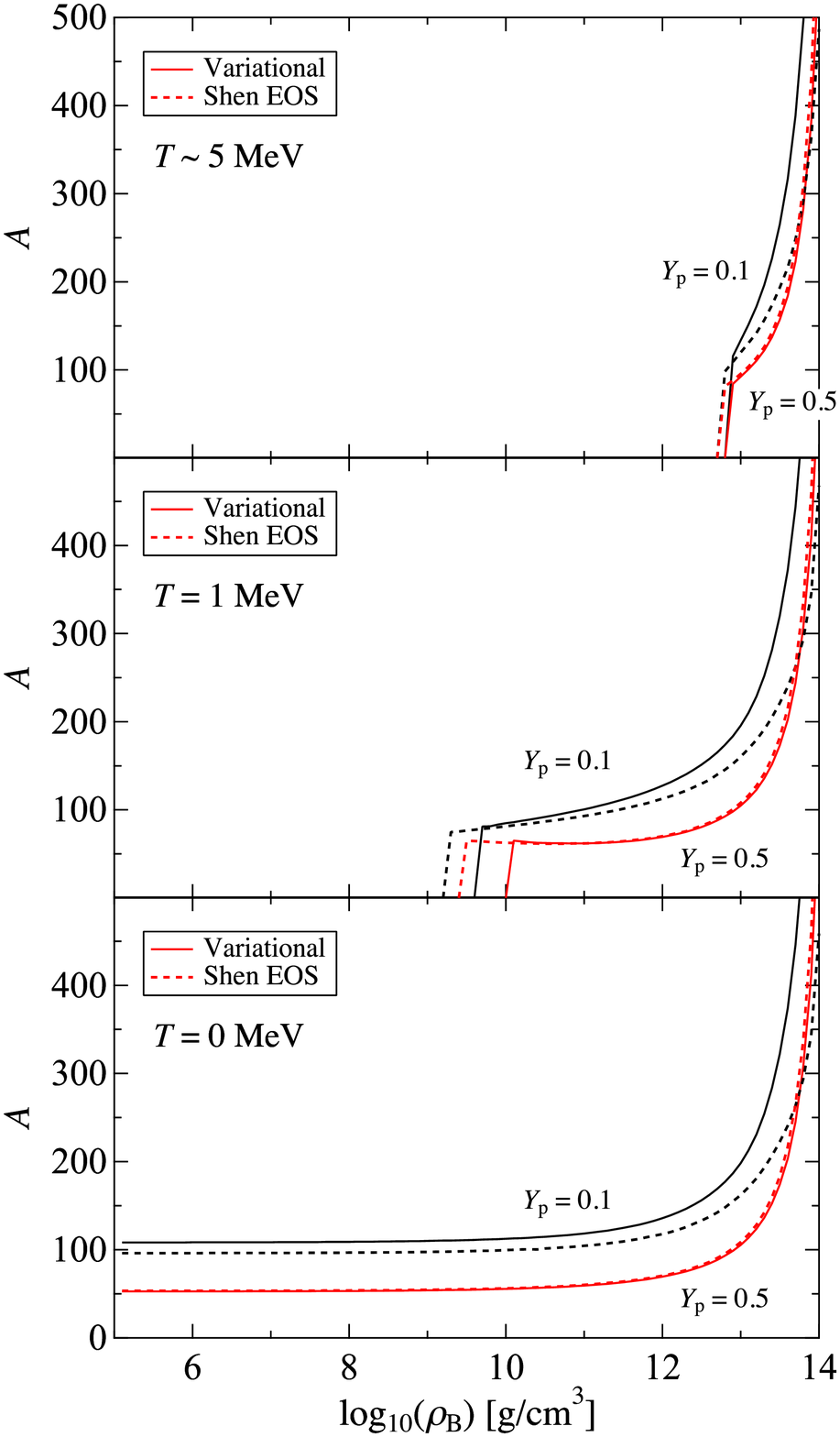}
\end{minipage} &
\begin{minipage}[t]{0.48\hsize}
\centering
\includegraphics[width=6.5cm]{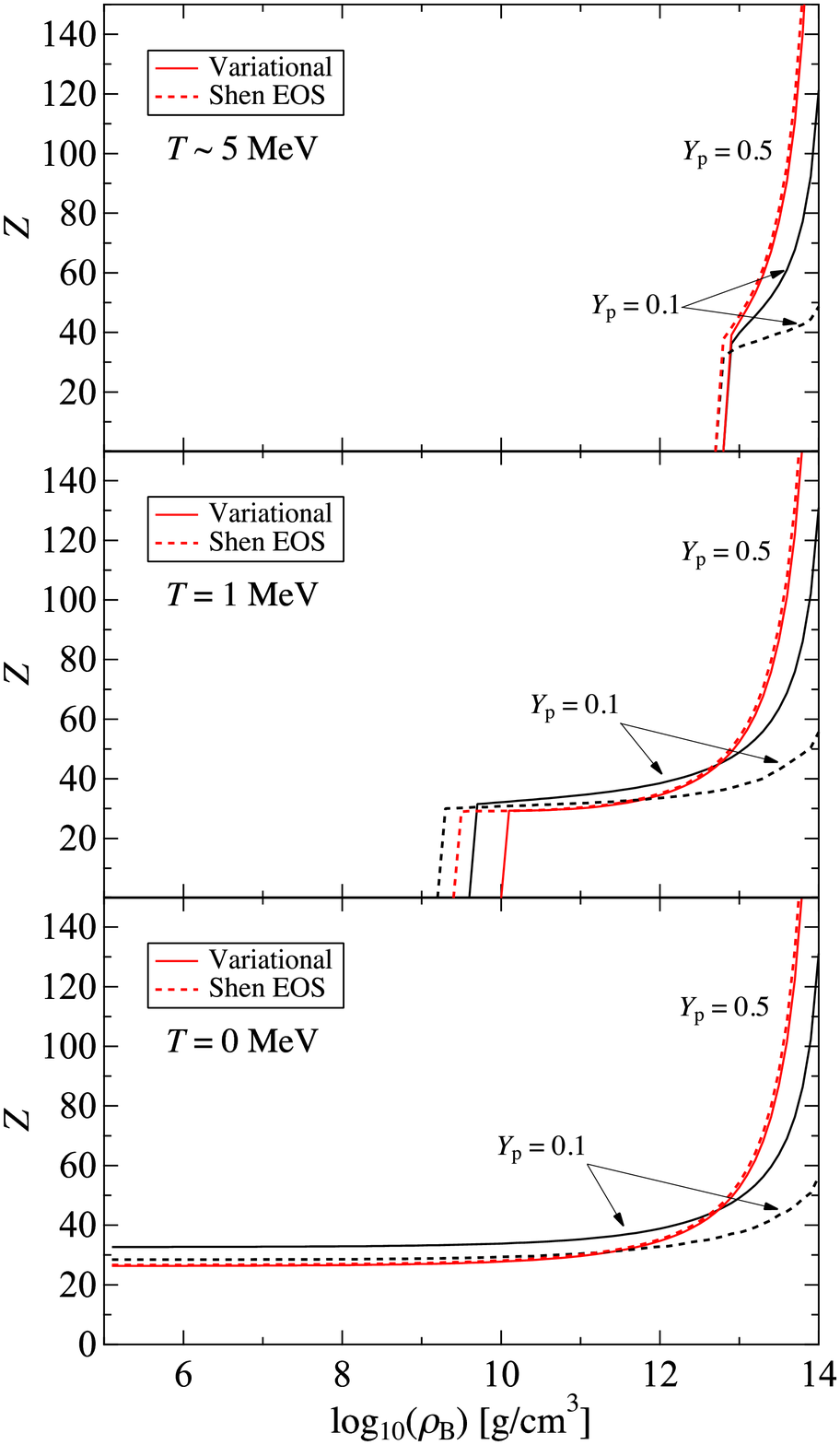}
\end{minipage} 
\end{tabular}
 \caption{(Color online) Mass number $A$ (left panels) and proton number $Z$ (right panels) as functions of the baryon mass density $\rho_{\mathrm{B}}$ 
 at the proton fractions $Y_{\mathrm{p}}$ = 0.1 and 0.5 at $T$ = 0, 1, and 5 MeV (from bottom to top panels). 
The solid and dashed lines are for our EOS and for the Shen EOS, respectively.}
\label{fig:AZ}
\end{figure}
\end{center}

\begin{figure}
  \centering
  \includegraphics[width=8.0cm]{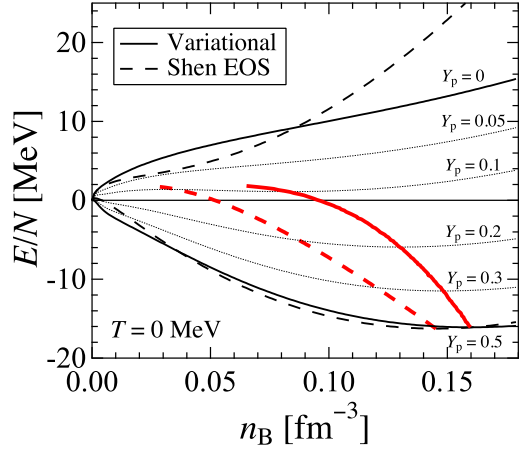}
  \caption{(Color online) Energies per nucleon of nuclear matter at zero temperature at various values of proton fraction $Y_{\mathrm{p}}$. 
The thin solid and dotted lines represent the results for our EOS, while the dashed lines are for the Shen EOS.
The thick solid and dashed lines connect the saturation points of asymmetric nuclear matter for our EOS and the Shen EOS, respectively.}
\label{fig:easym}
\end{figure}

\begin{figure}
  \centering
  \includegraphics[width=8.0cm]{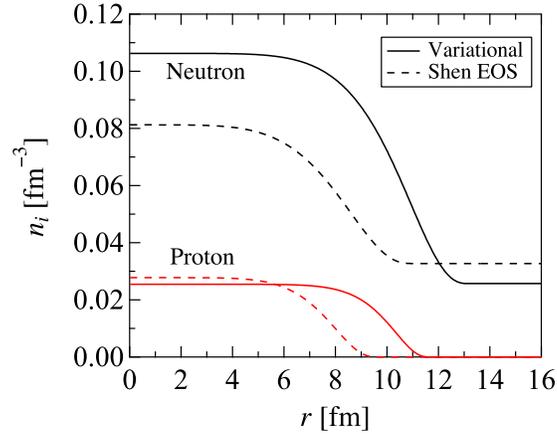}
  \caption{(Color online) Proton and neutron number density distributions in a Wigner-Seitz cell at $T=0$ MeV, $Y_{\mathrm{p}}=0.1$, and $\rho_{\mathrm{B}}=10^{13.9}\mathrm{g/cm}^3$.  
The solid and dashed lines are for our EOS and for the Shen EOS, respectively.}
\label{fig:distrb}
\end{figure}

The nuclear mass number $A$ and charge number $Z$ are shown in Fig.~\ref{fig:AZ} as functions of the baryon mass density $\rho_{\mathrm{B}}$ at $Y_{\mathrm{p}}$ = 0.1 and 0.5. 
It is seen that, in the case $Y_{\mathrm{p}}$ = 0.5, $A$ and $Z$ in our EOS are in good agreement with those in the Shen EOS. 
On the other hand, for neutron rich matter, $A$ and $Z$ in our EOS are larger than those in the Shen EOS.
This is for the following reason.
As shown in Fig.~\ref{fig:easym}, the saturation densities of asymmetric nuclear matter for our EOS are higher than those for the Shen EOS at zero temperature 
because the corresponding saturation density of symmetric nuclear matter for our EOS ($n_0$ = 0.16 fm$^{-3}$) is set higher than that for the Shen EOS ($n_0$ = 0.145 fm$^{-3}$).  
Another reason for our higher saturation density is the smaller value of $L$ for our EOS; as reported in Ref.~\cite{Oyak2}, 
EOSs with larger values of $L$ provide smaller saturation densities of asymmetric nuclear matter, even though the saturation densities of symmetric nuclear matter remain the same.  
Therefore, the central density of a nucleus, which is governed by the saturation density of asymmetric nuclear matter, is higher with our EOS. 
Furthermore, since the symmetry energy at subnuclear densities for our EOS is larger than that for the Shen EOS, 
corresponding to a smaller $L$ in our EOS, less neutrons drip out of a nucleus in a WS cell. 
Thus, as shown in Fig.~\ref{fig:distrb}, the difference in density between the nuclei and dripped neutrons is larger for our EOS. 
As a result, the density gradient at the nuclear surface, and corresponding surface energy become larger. 
Here we note that, in our Thomas-Fermi calculation, the gradient term $E_{\mathrm{grad}}$, 
which is balanced with the Coulomb energy through Eq.~(\ref{sizeeqlbr}), is about a half of the surface energy~\cite{Oyak1}. 
As the surface energy increases, therefore, the Coulomb energy, or corresponding $Z$, 
increases~\cite{Oyak2}, and our EOS with larger surface energy provides us a larger value of $Z$. 
Furthermore, the mass number $A$, which is associated with the proton number $Z$, also becomes larger than that of the Shen EOS.  

\begin{figure}
  \centering
  \includegraphics[width=8.0cm]{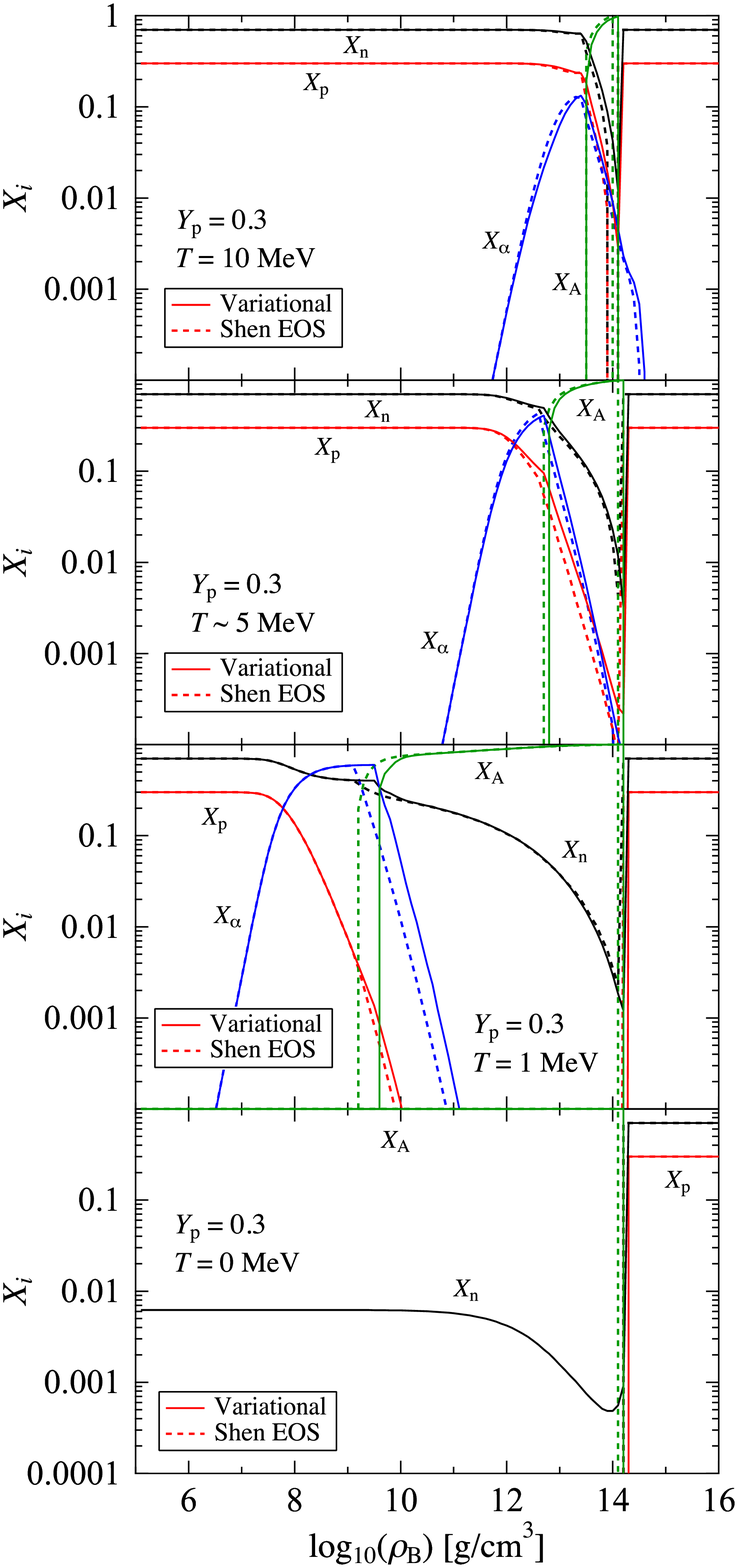}
 \caption{(Color online) Fractions of particles $X_i$ as functions of the baryon mass density $\rho_{\mathrm{B}}$ 
 at the proton fraction $Y_{\mathrm{p}}$ = 0.3 at $T$ = 0, 1, 5, and 10 MeV (from bottom to top panels).  
 The solid and dashed lines are for our EOS and for the Shen EOS, respectively.} 
\label{fig:frac}
\end{figure}

Figure \ref{fig:frac} shows fractions of neutrons, protons, alpha particles, and heavy nuclei $X_i$ ($i$ = $\mathrm{n}$, $\mathrm{p}$, $\alpha$, and $\mathrm{A}$) 
as functions of the baryon mass density $\rho_{\mathrm{B}}$ for $Y_{\mathrm{p}}$ = 0.3. 
While low-density nuclear matter at zero temperature is dominated by the heavy nuclei with dripped free neutrons, 
a homogeneous nucleon-gas phase appears in the low density region at finite temperatures. 
As $\rho_{\mathrm{B}}$ increases, alpha particles appear in hot nuclear matter and the alpha-particle fraction $X_{\alpha}$ increases. 
Then, after the heavy nuclei appear, $X_{\alpha}$ decreases rapidly because the formation of heavy nuclei exhausts most of the nucleons. 
As the temperature increases, the region of non-uniform phase becomes narrower and shifts to the higher density region.
It is also seen that the values of $X_i$ for our EOS are close to those for the Shen EOS. 

\begin{figure}
  \centering
  \includegraphics[width=8.0cm]{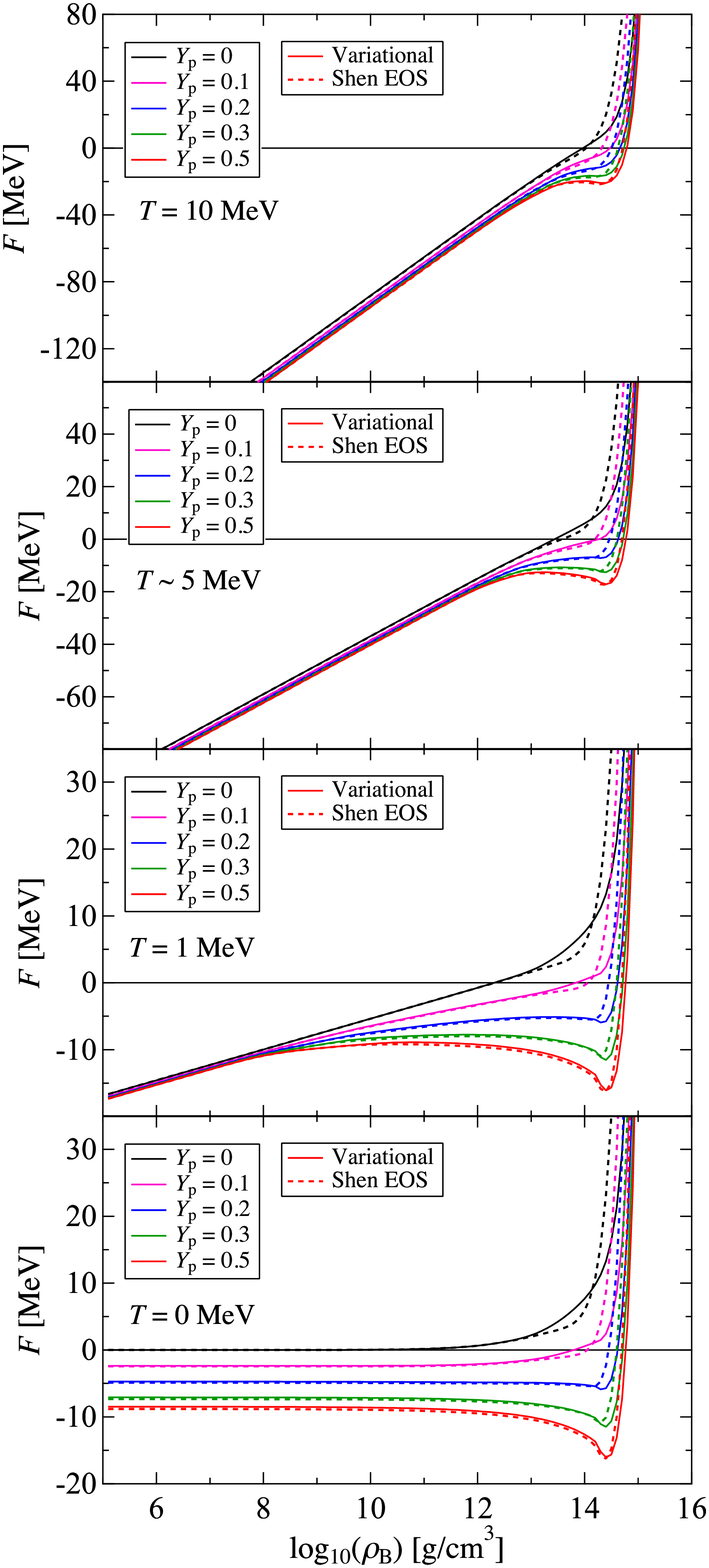}
  \caption{(Color online) Free energy per baryon as a function of the baryon mass density $\rho_{\mathrm{B}}$ at various values of $Y_{\mathrm{p}}$ at $T$ = 0, 1, 5, and 10 MeV (from bottom to top panels).
The solid and dashed lines are for our EOS and for the Shen EOS, respectively. 
In all the panels, the lines correspond, from top to bottom, to the cases at $Y_{\mathrm{p}}$ = 0, 0.1, 0.2, 0.3, and 0.5.}
\label{fig:F}
\end{figure}

We next show the free energy per baryon $F(n_{\mathrm{B}},Y_{\mathrm{p}},T)$ as a function of the baryon mass density $\rho_{\mathrm{B}}$ in Fig.~\ref{fig:F}. 
In the low-density region at finite temperatures, the free energy increases almost linearly with log$_{10}(\rho_{\mathrm{B}})$. 
However, as the density approaches 10$^{14}$ g/cm$^3$, this tendency is weakened and, for high values of $Y_{\mathrm{p}}$, $F(n_{\mathrm{B}},Y_{\mathrm{p}},T)$ decreases. 
This tendency is a result of the transition to a non-uniform phase.  
The free energies rise steeply owing to repulsive nuclear force at high densities, 
where large differences between our EOS and the Shen EOS can be seen reflecting the properties of uniform matter. 
In the homogeneous nucleon-gas phase at low densities and finite temperatures, 
our EOS is very close to the Shen EOS because the effect of interactions among nucleons is negligibly small. 

\begin{figure}
  \centering
  \includegraphics[width=8.0cm]{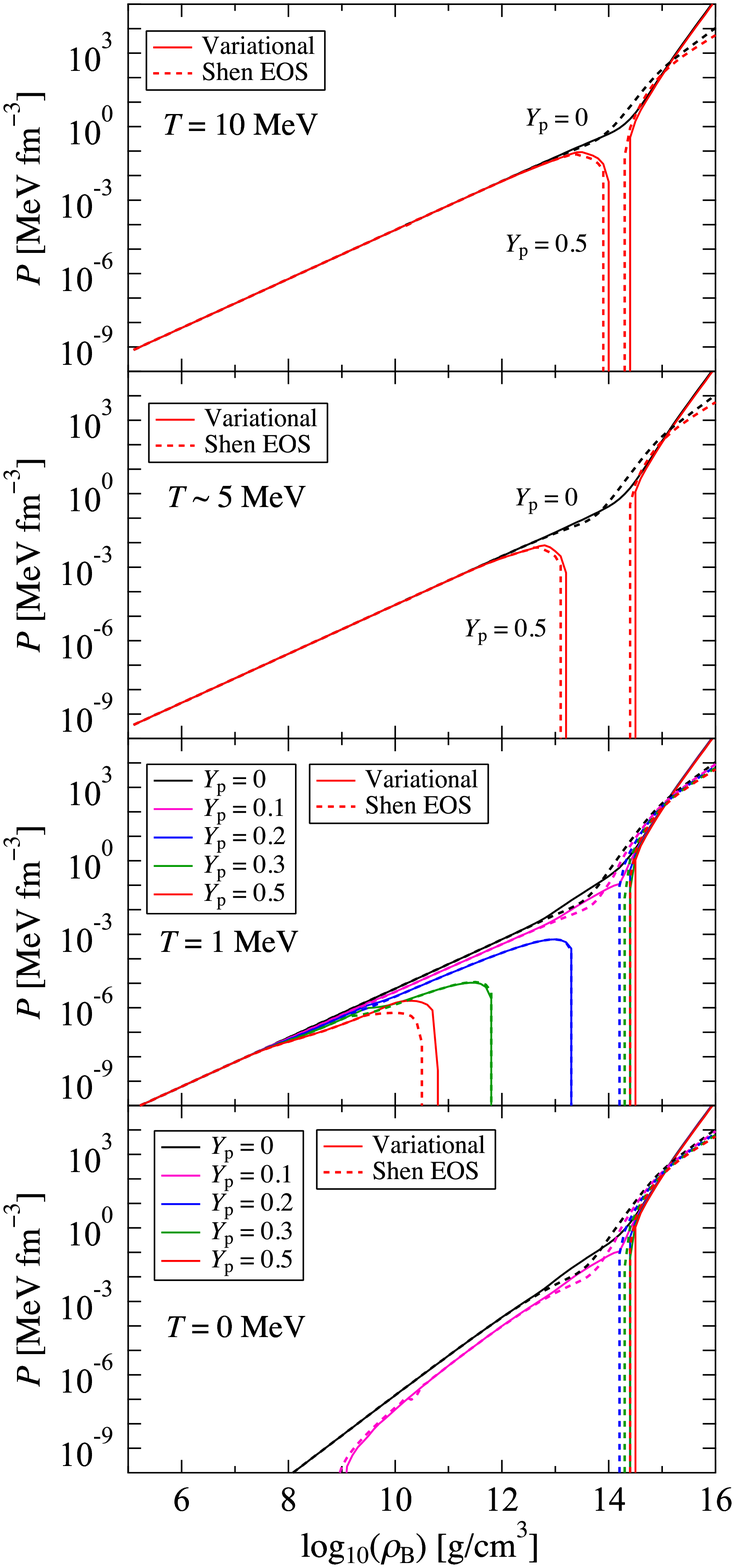}
  \caption{(Color online) Pressure as a function of the baryon mass density $\rho_{\mathrm{B}}$ at various values of $Y_{\mathrm{p}}$ at $T$ = 0, 1, 5, and 10 MeV (from bottom to top panels).
The solid and dashed lines are for our EOS and for the Shen EOS, respectively.
In the bottom two panels, the lines correspond, from top to bottom, to the cases at $Y_{\mathrm{p}}$ = 0, 0.1, 0.2, 0.3, and 0.5.}
\label{fig:P}
\end{figure}

Figure~\ref{fig:P} shows the pressure $P(n_{\mathrm{B}},Y_{\mathrm{p}},T)$ as a function of $\rho_{\mathrm{B}}$.  
For neutron-rich matter with smaller $Y_{\mathrm{p}}$, $P(n_{\mathrm{B}},Y_{\mathrm{p}},T)$ increases with $\rho_{\mathrm{B}}$ monotonically.  
In particular, the pressure is proportional to $\rho_{\mathrm{B}}$ at low densities at which the matter is regarded as a classical gas. 
For larger $Y_{\mathrm{p}}$, on the other hand, $P(n_{\mathrm{B}},Y_{\mathrm{p}},T)$ drops at the density where the formation of heavy nuclei occurs. 
We note that, in the realistic supernova core, the lepton and photon partial pressures are added, and then the total pressure increases monotonically with $\rho_{\mathrm{B}}$.   
The difference between $P(n_{\mathrm{B}},Y_{\mathrm{p}},T)$ of our EOS and that of the Shen EOS is not remarkable.  
At $Y_{\mathrm{p}}=0.2-0.3$, the densities at which $P(n_{\mathrm{B}},Y_{\mathrm{p}},T)$ rise steeply are higher than in the Shen EOS, 
because the transition density from the non-uniform phase to uniform phase is higher in our EOS, as discussed before.  

\begin{figure}
  \centering
  \includegraphics[width=8.0cm]{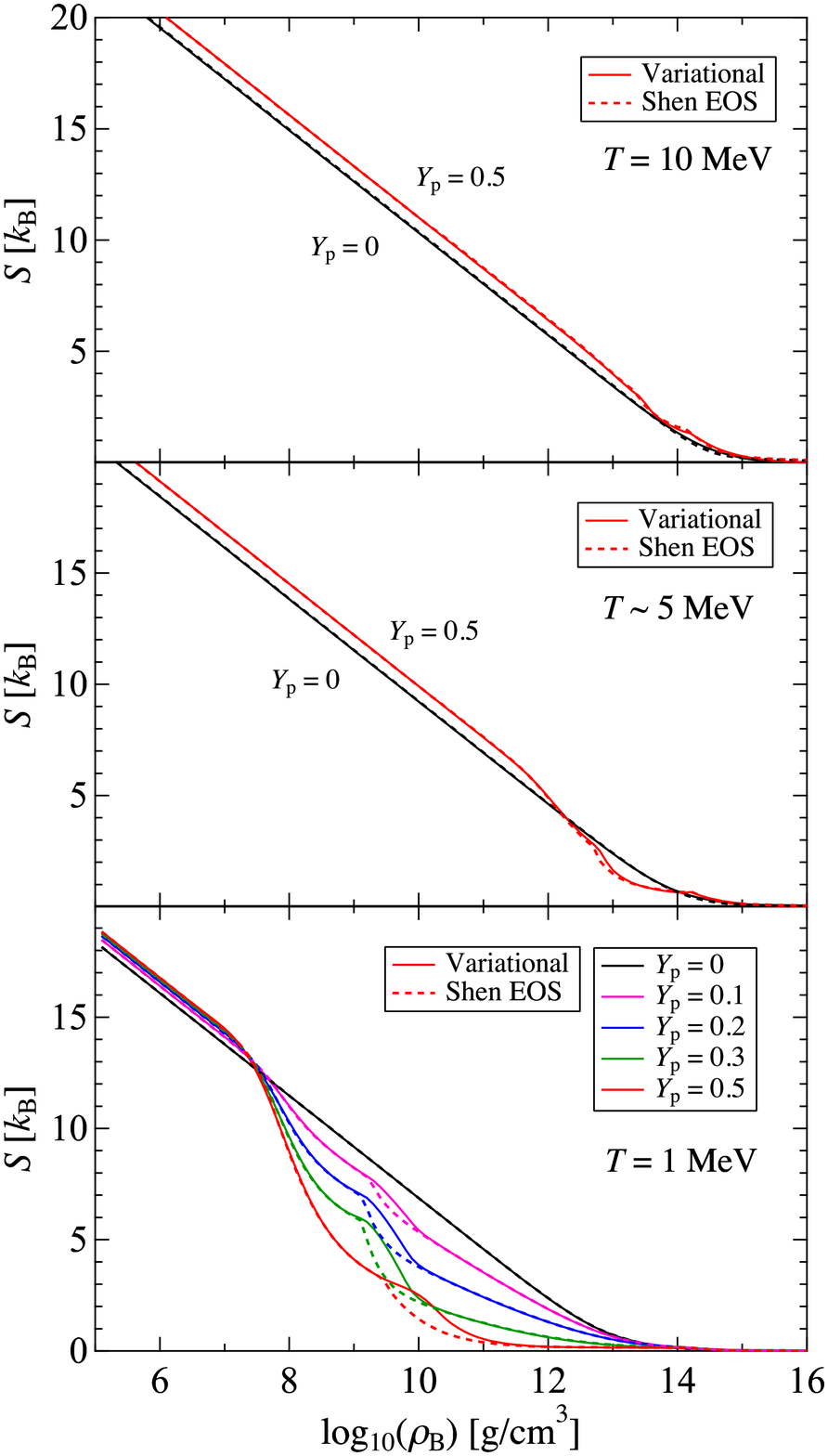}
  \caption{(Color online) Entropy per baryon as a function of the baryon mass density $\rho_{\mathrm{B}}$ at various values of $Y_{\mathrm{p}}$ at $T$ = 1, 5, and 10 MeV (from bottom to top panels). 
  The solid and dashed lines are for our EOS and for the Shen EOS, respectively. 
  In the bottom panel at $\rho_{\mathrm{B}}=10^9 \mathrm{g/cm^3}$, the lines correspond, from top to bottom, to the cases at $Y_{\mathrm{p}}$ = 0, 0.1, 0.2, 0.3, and 0.5.}
\label{fig:S}
\end{figure}

Figure~\ref{fig:S} shows the entropy per nucleon $S(n_{\mathrm{B}},Y_{\mathrm{p}},T)$ as a function of $\rho_{\mathrm{B}}$.  
It is seen that $S(n_{\mathrm{B}},Y_{\mathrm{p}},T)$ decreases as $\rho_{\mathrm{B}}$ increases.  
At $T=1$ MeV, $S(n_{\mathrm{B}},Y_{\mathrm{p}},T)$ drops at around $\rho_{\mathrm{B}} \simeq 10^8$g/cm$^3$ 
because of the mixing of  alpha particles, and it drops further around $\rho_{\mathrm{B}} \simeq 10^{10}$g/cm$^3$ owing to the formation of heavy nuclei.  
These drops are remarkable at higher $Y_{\mathrm{p}}$ because both the mixing of alpha particles and the formation of heavy nuclei occur more easily at higher $Y_{\mathrm{p}}$.  
Owing to the formation of heavy nuclei, $S(n_{\mathrm{B}},Y_{\mathrm{p}},T)$ decreases as $Y_{\mathrm{p}}$ increases in the non-uniform phase: 
In contrast, $S(n_{\mathrm{B}},Y_{\mathrm{p}},T)$ increases with $Y_{\mathrm{p}}$ in the uniform phase.  
There is no significant difference between $S(n_{\mathrm{B}},Y_{\mathrm{p}},T)$ of our EOS and that of the Shen EOS, 
although a slight variance is seen at $\rho_{\mathrm{B}} \simeq 10^{9}$g/cm$^3$, which corresponds to the difference in the onset density of heavy nuclei formation.
As $T$ increases, $S(n_{\mathrm{B}},Y_{\mathrm{p}},T)$ increases, and the drop of $S(n_{\mathrm{B}},Y_{\mathrm{p}},T)$ in the non-uniform density region is weakened.  

\section{Summary}
In this paper, we constructed a new nuclear EOS for core collapse supernova simulations using the cluster variational method and the Thomas-Fermi approximation.  
For uniform nuclear matter, as reported in Ref.~\cite{T1}, we calculated the free energy per nucleon with the cluster variational method starting from the AV18 two-nucleon potential and the UIX three-nucleon potential.  
The obtained (free) energies per nucleon at zero and finite temperatures are consistent with the results by APR and by Mukhargee with the FHNC variational calculations.  
To apply the obtained free energy of uniform nuclear matter to the Thomas-Fermi calculations for non-uniform matter, 
we removed the formation of deuteron clusters in nuclear matter at low densities by modifying the healing-distance condition.  
In this process, we reproduced the virial EOS of neutron matter at finite temperatures as well as the neutron-matter EOS at zero temperature with the GFMC calculations.
Using this free energy of uniform nuclear matter, we performed the Thomas-Fermi calculations for non-uniform nuclear matter.  
Furthermore, we took into account the mixing of alpha particles in uniform and non-uniform nuclear matter as in the case of the Shen EOS.  

At low $Y_{\mathrm{p}}$, the critical density with respect to the phase transition from non-uniform matter to uniform matter is higher than that of the Shen EOS.  
It is also found that the critical temperature decreases more moderately with $Y_{\mathrm{p}}$ for our EOS.  
Furthermore, the mass numbers $A$ and proton numbers $Z$ of nuclei appearing in the non-uniform phase are larger than in the Shen EOS at low $Y_{\mathrm{p}}$.  
These results are understood by recognizing that the density derivative coefficient of the symmetry energy $L$ of our EOS is smaller than that of the Shen EOS; 
the last result is also affected by the difference between the saturation densities of symmetric nuclear matter of these two EOSs.  

To our knowledge, this is the first nuclear EOS for core collapse supernova simulations based on realistic nuclear forces; 
this EOS enables us to study the mechanism of core-collapse supernovae in terms of nuclear forces.  

Since this EOS reproduces the observationally suggested radii of neutron stars, it is also interesting to apply it to other high-energy astrophysical phenomena such as neutron star mergers.  
Furthermore, it will be rewarding to study hyperon mixing in nuclear matter with our EOS.  
In fact, we have already extended this cluster variational method for nuclear matter at zero temperature to hyperonic nuclear matter containing $\Lambda$ and $\Sigma^-$ hyperons \cite{T3}. 
An extension of the variational calculation to hot hyperonic nuclear matter with the goal of developing a new supernova EOS table with hyperons is in progress. 

\section*{Acknowledgements}
We would like to express special thanks to H. Kanzawa, K. Oyamatsu, K. Sumiyoshi, M. Yamada and S. Yamada for valuable discussions and comments, 
to H.~Shen for providing us with numerical data on parameters for the Thomas-Fermi calculation, 
and to H. Matsufuru for supporting on parallel computing. 
Early stages of the healing-distance modification were performed by Y. Nagano. 
The numerical computations in this work were carried out on SR16000 at YITP in Kyoto University, 
on SR16000 and Blue Gene/Q at the High Energy Accelerator Research Organization (KEK), and on SR16000 at the Information Technology Center of the University of Tokyo.
This work is supported by JSPS (Nos. 22540296, 23224006, 24244036, 25400275, 26870615), 
the Grant-in-Aid for Scientific Research on Innovative Areas of MEXT (Nos. 20105003, 20105004, 20105005, 24105008, 26104006, 26105515), 
RIKEN iTHES Project, and Special Postdoctoral Researcher Program of RIKEN. 

\appendix
\section{Contents in the table of the supernova equation of state}
The EOS table is available on the Web at \\ {\tt http://www.np.phys.waseda.ac.jp/EOS/}.  
For convenience, the ranges and grids of $T$, $Y_{\mathrm{p}}$, and $\rho_{\mathrm{B}}$ are the same as those of the Shen EOS, as shown in Table~\ref{table1}. 
In contrast with the case of the Shen EOS, we do not tabulate the values of $m_i^*$ included in Eq.~(\ref{eq:nk}), 
because further careful studies would be necessary to clarify the physical meaning of $m_i^*$ appearing in our variational calculation. 
Except for the effective mass, the physical quantities listed in the supernova EOS table are consistent with those in the Shen EOS, as follows:

\begin {table}[b]
\caption{Ranges of temperature $T$, proton fraction $Y_{\mathrm{p}}$, and baryon mass density $\rho_{\mathrm{B}}$ in the table of the variational EOS. 
At the top of the last column, "$+1$" represents the case at $T=0$ MeV.}
\label{table1}
\begin {center}
\begin{tabular}{ccccc} \hline
   Parameter & Minimum & Maximum & Mesh & Number \\ \hline
   $\mathrm{log}_{10} (T)$ [MeV] & $-1.00$ & $2.60$ & $0.04$ & $91+1$   \\  
   $Y_{\mathrm{p}} $ & $0$ & $0.65$ & $0.01$ & $66$   \\  
    $\mathrm{log}_{10} (\rho_{\mathrm{B}})$ [g/cm$^3$] & $5.1$ & $16.0$ & $0.1$ & $110$   \\    \hline
\end{tabular}
\end {center}
\end{table}

\begin{enumerate}	
	\item Logarithm of baryon mass density: log$_{10}(\rho_{\mathrm{B}})[\mathrm{g~cm}^{-3}]$. 
	
	\item Baryon number density: $n_{\mathrm{B}}[\mathrm{fm}^{-3}]$. 

	The baryon number density $n_{\mathrm{B}}$ is defined as in Eqs.~(\ref{nBuni}) and (\ref{nBnonuni}) for uniform and non-uniform nuclear matter, respectively. 
	The baryon number density is related to the baryon mass density as $\rho_{\mathrm{B}}=m_{\mathrm{u}}n_{\mathrm{B}}$ with $m_{\mathrm{u}}c^2=931.494$ MeV, 
	following the definition in the Shen EOS. 
	
	\item Proton fraction: $Y_{\mathrm{p}}$. 
	
	The proton fraction $Y_{\mathrm{p}}$ is defined as in Eqs.~(\ref{Ypuni}) and (\ref{Ypnonuni}) for uniform and non-uniform nuclear matter, respectively. 

	\item Free energy per nucleon: $F$[MeV]. 
	
	The free energy per nucleon $F$ is defined as in Eqs.~(\ref{Funi}) and (\ref{Fnonuni}) for uniform and non-uniform nuclear matter, respectively.
	The nucleon rest mass energy is not included in $F$. This definition is consistent with that in the Shen EOS. 
	The free energy per nucleon with the rest mass energy is given by 
	\begin{equation}
	{\cal F} = F + Y_{\mathrm{p}}m_{\mathrm{p}}c^2 + (1-Y_{\mathrm{p}})m_{\mathrm{n}}c^2, 
	\end{equation}
	with $m_{\mathrm{p}}c^2=938.272$ MeV and $m_{\mathrm{n}}c^2=939.565$ MeV. 

	\item Internal energy per nucleon: $E_{\mathrm{int}}$[MeV].
	
	The internal energy per nucleon relative to the atomic mass unit, $E_{\mathrm{int}}$, is defined as in Eq.~(\ref{themEint}).

	\item Entropy per nucleon: $S$[$k_{\mathrm{B}}$]. 
	
	The entropy per nucleon $S$ is defined as in Eq.~(\ref{themS}). 
	
	\item Mass number of the heavy nucleus: $A$.
	
	The mass number of the heavy nucleus $A$ is defined as in Eq.~(\ref{defA}). 
	
	\item Proton number of the heavy nucleus: $Z$. 
	
	The proton number of the heavy nucleus $Z$ is defined as in Eq.~(\ref{defZ}). 
	
	\item Free neutron fraction: $X_\mathrm{n}$.
	
	The free neutron fraction $X_\mathrm{n}$ is defined as in Eq.~(\ref{XiTF}).
	
	\item Free proton fraction: $X_\mathrm{p}$.
	
	The free proton fraction $X_\mathrm{p}$ is defined as in Eq.~(\ref{XiTF}). 
	
	\item Free alpha-particle fraction: $X_{\alpha}$.
	
	The free alpha-particle fraction $X_{\alpha}$ is defined as in Eq.~(\ref{XalphaTF}). 
	
	\item Heavy nucleus fraction: $X_{\mathrm{A}}$.
	
	The heavy nucleus fraction $X_{\mathrm{A}}$ is defined as in Eq.~(\ref{XA}). 
	
	\item Pressure: $P$ [MeV fm$^{-3}$].
	
	The pressure $P$ is defined as in Eq.~(\ref{themP}). 
		
	\item Neutron chemical potential: $\mu_{\mathrm{n}}$ [MeV].
	 
	The neutron chemical potential relative to the neutron rest mass energy, $\mu_{\mathrm{n}}$, is defined as in Eq.~(\ref{themMUn}). 
			
	\item Proton chemical potential: $\mu_{\mathrm{p}}$ [MeV].
	
	The proton chemical potential relative to the proton rest mass energy, $\mu_{\mathrm{p}}$, is defined as in Eq.~(\ref{themMUp}). 
\end{enumerate}

\section{Detailed calculation procedure}
In this appendix, we present the numerical calculation procedure for the construction of the supernova EOS in detail. 

\subsection {Uniform nuclear matter}
For uniform nuclear matter (Section 2.2), we tabulate $f_{\mathrm{N}}(n_{\mathrm{p}}, n_{\mathrm{n}},T)$ in Eq.~(\ref{Funi0}) in the ranges and grids shown in Table~\ref{table2} for the use of the subsequent numerical calculations, i.e., the mixing of alpha particles and the Thomas-Fermi approximation. 
Hereinafter, this input table for $f_{\mathrm{N}}(n_{\mathrm{p}}, n_{\mathrm{n}},T)$ is referred to as table I. 
In addition to table I, we also tabulate the free energy of uniform nuclear matter $F_{\mathrm{N}}(n_{\mathrm{p}}, n_{\mathrm{n}},T)$ in the wide ranges of temperature $T$, proton fraction $Y_{\mathrm{p}}$, and baryon mass density $\rho_{\mathrm{B}}$ shown in Table~\ref{table3}.  
Hereinafter, this table for $F_{\mathrm{N}}(n_{\mathrm{p}}, n_{\mathrm{n}},T)$ is referred to as table II, 
which will be used in the calculation of the alpha-particle mixing for uniform nuclear phase in the next step.

\begin {table}[b]
\caption{
Ranges of temperature $T$, neutron excess squared $\zeta$, and baryon number density $n_{\mathrm{B}}$ in the table of the free energy density of uniform nuclear matter $f_{\mathrm{N}}(n_{\mathrm{p}}, n_{\mathrm{n}},T)$, (Table I). At the top of the last column, "$+1$" represents the case at $T=0$ MeV.}
\label{table2}
\begin {center}
\begin{tabular}{ccccc} \hline
   Parameter & Minimum & Maximum & Mesh & Number \\ \hline
   $\mathrm{log}_{10} (T)$ [MeV] & $-1.08$ & $1.52$ & $0.04$ & $66+1$   \\  \hline
   $\zeta = [(n_{\mathrm{n}}-n_{\mathrm{p}})/n_{\mathrm{B}}]^2$ & $0$ & $0.8$ & $0.00625$ & $129$   \\
       & $0.803125$ & $0.95$ & $0.003125$ & $48$   \\
       & $0.9515625$ & $0.99375$ & $0.0015625$ & $28$   \\
       & $0.99453125$ & $1$ & $0.00078125$ & $8$   \\  \hline
    $n_{\mathrm{B}}$ [fm$^{-3}$] & $0.000001$ & $0.0001$ & $0.000001$ & $100$   \\  
       & $0.00011$ & $0.001$ & $0.00001$ & $90$   \\
       & $0.0011$ & $0.18$ & $0.0001$ & $1790$   \\  \hline
\end{tabular}
\end {center}
\end{table}

\begin {table}[b]
\caption{
Ranges of temperature $T$, proton fraction $Y_{\mathrm{p}}$, and baryon mass density $\rho_{\mathrm{B}}$ in the table of the free energy per nucleon (Tables II and III). 
At the top of the last column, "$+1$" represents the case at $T=0$ MeV.}
\label{table3}
\begin {center}
\begin{tabular}{ccccc} \hline
   Parameter & Minimum & Maximum & Mesh & Number \\ \hline
   $\mathrm{log}_{10} (T)$ [MeV] & $-1.08$ & $2.68$ & $0.04$ & $95+1$   \\  \hline
   $Y_{\mathrm{p}}$ & $0$ & $0.02$ & $0.005$ & $5$   \\  
   & $0.03$ & $0.67$ & $0.01$ & $65$   \\  \hline
    $\mathrm{log}_{10} (\rho_{\mathrm{B}})$ [g/cm$^3$] & $4.9$ & $16.2$ & $0.1$ & $114$   \\    \hline
\end{tabular}
\end {center}
\end{table}

\subsection {Alpha-particle mixing in uniform nuclear phase}
For the calculation of the alpha-particle mixing in uniform nuclear phase (Section 2.3), we minimize $f(n_{\mathrm{p}}, n_{\mathrm{n}}, n_{\alpha},T)$ in Eq.~(\ref{Fbulk2}) numerically with respect to $n_{\alpha}$ for a fixed set ($T$, $n_{\mathrm{B}}$, $Y_{\mathrm{p}}$) to determine the optimal $n_{\alpha}$. 
In this minimization, the free energy contribution from nucleons $f_{\mathrm{N}}(\tilde{n}_{\mathrm{p}}, \tilde{n}_{\mathrm{n}},T)$ in Eq.~(\ref{Fbulk2}) is calculated from the data in table I by the interpolation with respect to the neutron excess squared $\zeta = [(n_{\mathrm{n}}-n_{\mathrm{p}})/n_{\mathrm{B}}]^2$: 
As reported in Ref.~\cite{T1}, the symmetry free energy of uniform nuclear matter is roughly proportional to $\zeta$. 
When the effective nucleon number density becomes lower than the lowest density in table I, i.e., $\tilde{n}_{\mathrm{p}} + \tilde{n}_{\mathrm{n}} < 10^{-6}$ fm$^{-3}$, 
we adopt the following ideal-gas approximation for $f_{\mathrm{N}}(\tilde{n}_{\mathrm{p}}, \tilde{n}_{\mathrm{n}},T)$; 
\begin{equation}
f_{\mathrm{N}}(\tilde{n}_{\mathrm{p}}, \tilde{n}_{\mathrm{n}},T) = 
-T \tilde{n}_{\mathrm{p}} \left[ \ln \frac{2n_{\mathrm{Q}}}{\tilde{n}_{\mathrm{p}}} +1 \right] 
-T \tilde{n}_{\mathrm{n}} \left[ \ln \frac{2n_{\mathrm{Q}}}{\tilde{n}_{\mathrm{n}}} +1 \right]. 
\label{Boltzmann}
\end{equation}
In addition, we employ the data in table II when $\tilde{n}_{\mathrm{p}} + \tilde{n}_{\mathrm{n}} > 0.18$ fm$^{-3}$ or $\mathrm{log}_{10} (T) > 1.52$ MeV, which are not covered in table I. 
Finally, using the determined $n_{\alpha}$, we recalculate $f_{\mathrm{N}}(\tilde{n}_{\mathrm{p}}, \tilde{n}_{\mathrm{n}},T)$ with the cluster variational method reported in Section 2.2 
to obtain the precise minimal free energy density $f(n_{\mathrm{p}}, n_{\mathrm{n}}, n_{\alpha},T)$. 

We then tabulate the free energy per baryon including alpha particles $F(n_{\mathrm{B}}, Y_{\mathrm{p}},T)$ in Eq.~(\ref{Funi}) in ranges and grids as shown in Table~\ref{table3}. 
Hereinafter, this table is referred to as table III. 
In this table, we take into account the alpha-particle mixing when the alpha-particle fraction $X_{\alpha} \equiv 4n_{\alpha}/n_{\mathrm{B}} \geq 10^{-15}$, 
while, for $X_{\alpha} < 10^{-15}$ or at zero temperature, $F_{\mathrm{N}}(n_{\mathrm{p}}, n_{\mathrm{n}},T)$ in table II is adopted as $F(n_{\mathrm{B}}, Y_{\mathrm{p}},T)$ in table III.
Then, for  $10^{-50} < X_{\alpha} < 10^{-15}$, we evaluate $n_{\alpha}$ with the following expression for the non-interacting Boltzmann gas: 
\begin{equation}
n_{\alpha} = 8n_{\mathrm{Q}} \exp\left(\frac{2\mu_{\mathrm{n}}+2\mu_{\mathrm{p}}+B_{\alpha}}{T} \right). \label{dripMUalpha}
\end{equation}
Here, the chemical potentials of protons $\mu_{\mathrm{p}}(n_{\mathrm{B}}, Y_{\mathrm{p}},T)$ and neutrons $\mu_{\mathrm{n}}(n_{\mathrm{B}}, Y_{\mathrm{p}},T)$ are obtained by Eqs.~(\ref{themMUp}) and (\ref{themMUn}). 

\subsection {Thomas-Fermi calculation}
In the Thomas-Fermi calculation (Section 3), we minimize the free energy density $F_{\mathrm{cell}}/V_{\mathrm{cell}}$ numerically with respect to eight independent parameters among $a$, $n_{\mathrm{n}}^{\mathrm{in}}$, $n_{\mathrm{n}}^{\mathrm{out}}$, $R_{\mathrm{n}}$, $t_{\mathrm{n}}$, $n_{\mathrm{p}}^{\mathrm{in}}$, $n_{\mathrm{p}}^{\mathrm{out}}$, $R_{\mathrm{p}}$, $t_{\mathrm{p}}$, and $n_{\alpha}^{\mathrm{out}}$ for each $n_{\mathrm{B}}$, $Y_{\mathrm{p}}$ and $T$. 

For the bulk free energy $F_{\mathrm{bulk}}$ in Eq.~(\ref{Fbulk}), $f(n_{\mathrm{p}}(r), n_{\mathrm{n}}(r), n_{\alpha}(r),T)$ is calculated from table I and the ideal-gas approximation in Eq.~(\ref{Boltzmann}), as in Appendix B.2. 
At zero temperature, since the ideal-gas approximation is not available, we use a function $C[n_{\mathrm{p}}^{5/3}+n_{\mathrm{n}}^{5/3}]$ based on the Fermi-gas model. 
Here, the coefficient $C$ is chosen so that $C[n_{\mathrm{p}}^{5/3}+n_{\mathrm{n}}^{5/3}]$ smoothly connects to data in table I 
at $n_{\mathrm{p}} + n_{\mathrm{n}} = 10^{-6}$ fm$^{-3}$ for each proton fraction. 
The gradient term $E_{\mathrm{grad}}$ and Coulomb energy term $E_{\mathrm{C}}$ in Eq.~(\ref{Fcell}) are expressed analytically with the above parameters. 

We then perform the multidimensional minimization of $F_{\mathrm{cell}}/V_{\mathrm{cell}}$ with the simplex method. 
In this minimization, it is difficult to obtain converging solutions accurately in the Thomas-Fermi calculation when the value of $n_i^{\mathrm{out}}$ ($i=$ p, n, or $\alpha$) becomes extremely small. 
Thus, if $X_i$ ($i=$ p, n, or $\alpha$) is estimated to be smaller than $10^{-10}$ by the Thomas-Fermi calculation, we evaluate the value of $n_{i}^{\mathrm{out}}$ from the following Boltzmann-gas expressions  
\begin{equation}
{n_{j}^{\mathrm{out}} = 2n_{\mathrm{Q}} \exp\left(\frac{\mu_{j}}{T} \right)} \  \ (j= \mathrm{p, n})\label{dripMUpn} 
\end{equation}
or Eq.~(\ref{dripMUalpha}) for $n_{\alpha}^{\mathrm{out}}$, instead of the Thomas-Fermi calculation. 
Here, the chemical potentials are obtained by Eqs.~(\ref{themMUp}) and (\ref{themMUn}). 
We then recalculate $F(n_{\mathrm{B}}, Y_{\mathrm{p}},T)$ using the determined parameter set ($a$, $n_{\mathrm{n}}^{\mathrm{in}}$, $n_{\mathrm{n}}^{\mathrm{out}}$, $R_{\mathrm{n}}$, $t_{\mathrm{n}}$, $n_{\mathrm{p}}^{\mathrm{in}}$, $n_{\mathrm{p}}^{\mathrm{out}}$, $R_{\mathrm{p}}$, $t_{\mathrm{p}}$, $n_{\alpha}^{\mathrm{out}}$) for each $n_{\mathrm{B}}$, $Y_{\mathrm{p}}$ and $T$, 
and tabulate the obtained $F(n_{\mathrm{B}}, Y_{\mathrm{p}},T)$ in ranges and grids as shown in Table~\ref{table3}. 

By comparing this table for non-uniform nuclear matter with table III completed in Appendix B.2, we finally choose the optimal value of the free energy $F(n_{\mathrm{B}}, Y_{\mathrm{p}},T)$ for each $n_{\mathrm{B}}$, $Y_{\mathrm{p}}$ and $T$.  
The obtained set of $F(n_{\mathrm{B}}, Y_{\mathrm{p}},T)$ covers the whole ranges and grids shown in Table~\ref{table3}. 
Using this complete set of $F(n_{\mathrm{B}}, Y_{\mathrm{p}},T)$, 
other thermodynamic quantities are calculated through the thermodynamic relations~(\ref{themP})-(\ref{themMUn}), 
and the resultant EOS is tabulated in ranges and grids as shown in Table~\ref{table1}.
The partial derivatives are performed numerically with a five-point differentiation method. 


\end{document}